\documentclass{emulateapj}

\usepackage{graphicx}

\slugcomment{The Astrophysical Journal, In Press}

\shorttitle{UNSTABLE DISK GALAXIES. II.}
\shortauthors{M. A. Jalali}

\begin{document}

\title{UNSTABLE DISK GALAXIES. II. THE ORIGIN OF GROWING AND STATIONARY MODES}

\author{Mir Abbas Jalali
}
\affil{Sharif University of Technology, Azadi Avenue, Tehran, Iran;
mjalali@sharif.edu \\
Institute for Advanced Study, Einstein Drive, Princeton, NJ 08540}


\begin{abstract}

I decompose the unstable growing modes of stellar disks to 
their Fourier components and present the physical mechanism 
of instabilities in the context of resonances. When the equilibrium
distribution function is a non-uniform function of the orbital
angular momentum, the capture of stars into the corotation 
resonance imbalances the disk angular momentum and triggers 
growing bar and spiral modes. The stellar disk can then recover 
its angular momentum balance through the response of non-resonant 
stars. I carry out a complete analysis of orbital structure 
corresponding to each Fourier component in the radial angle,
and present a mathematical condition for the occurrence of 
van Kampen modes, which constitute a continuous family. 
I discuss on the discreteness and allowable pattern speeds 
of unstable modes and argue that the mode growth is saturated 
due to the resonance overlapping mechanism. An individually 
growing mode can also be suppressed if the corotation and 
inner Lindblad resonances coexist and compete to capture a 
group of stars. Based on this mechanism, 
I show that self-consistent scale-free disks with a sufficient 
distribution of non-circular orbits should be stable under 
perturbations of angular wavenumber $m>1$. I also derive a 
criterion for the stability of stellar disks against 
non-axisymmetric excitations.  
\end{abstract}

\keywords{stellar dynamics,
          instabilities,
          methods: analytical,
          galaxies: kinematics and dynamics,
          galaxies: spiral,
          galaxies: structure}

\section{INTRODUCTION}
\label{sec:intro}

Both $N$-body simulations \citep{H71} and analytical methods
(Kalnajs 1978; Jalali \& Hunter 2005, hereafter JH) show that global
instabilities can generate barred structures in galactic disks.
Apart from the bar mode, which is an isolated event in frequency
space, global spiral modes seen in the eigenspectra of cored stellar
disks (Jalali 2007, hereafter Paper I) constitute a discrete family
that bifurcates form stationary \citet{vK55} modes. But not all
spiral structures are global modes as disturbances induced by close 
neighbors \citep{BH92} and density inhomogeneities \citep{T90} may 
also create the spiral patterns of the observed galaxies.

A mode of a stellar disk is a mathematical entity that comes out of
an eigenvalue problem. However, its physical origin in isolated
systems has not yet been understood clearly. Lynden-Bell \& Kalnajs
(1972, hereafter LBK) attempted to explain a mode through the
transport of angular momentum between different parts of the disk.
They suggested that the inner Lindblad resonance (ILR) releases the
angular momentum of central regions and the spiral structure
transports it to the outer parts through the corotation (CR) and
outer Lindblad resonances (OLR). This mechanism is favored by some
galactic dynamicists \citep{ATHA03}, but it is seriously challenged
by Toomre's (1981) theory that says that feedback through the
galactic center is a critical ingredient for growing modes. In
Toomre's theory, on the other hand, the modeling of feedback as the
reflection of a leading spiral wave at the galactic center and its
emergence as a trailing one, is a simple description of a very
complicated dynamics that governs the motions of stars. Unresolved
issues concerning the evolution of unstable modes include the
following: (i) The swing amplification theory is not capable of 
predicting the fate of a growing mode against other stationary 
and unstable modes that coexist in the eigenspectrum of a given 
model. (ii) How does the bar mode saturate? \citep{KJKJ07}.
(iii) Why does the nonlinear bar terminate almost at the corotation 
radius? \citep{S81} (iv) We should also understand the origin of 
different species in an eigenspectrum and interpret their continuous 
or discrete nature, and the distribution of their pattern speeds and
growth rates.

Stellar orbits in a galactic disk begin to evolve once the surface 
density deviates from its equilibrium state, 
$\Sigma_0(\textbf{\textit{x}})$, and develops a time-varying 
mean-field potential $V_1(\textbf{\textit{x}},t)$. Here 
$\textbf{\textit{x}}$ denotes the position vector of stars at the 
time $t$. In the linear regime, we are usually interested in density 
waves that grow/decay according to the exponential law $e^{st}$ 
and rotate with the fixed pattern speed $\Omega_p$. For a wave 
of $m$-fold symmetry, the perturbed potential becomes
\begin{equation}
V_1=\epsilon e^{st} \tilde V 
\left (\textbf{\textit{x}},m\Omega_p t \right ).
\label{eq:perturbed-potnetial-introduction}
\end{equation}
Since we are dealing with infinitesimal perturbations, 
I have introduced the small parameter $\epsilon$ so that 
$\epsilon e^{st}\ll 1$. 
From (\ref{eq:perturbed-potnetial-introduction}) one 
arrives at the equations of motion
\begin{equation}
\dot \textbf{\textit{x}} = \textbf{\textit{v}},~~
\dot \textbf{\textit{v}} = 
-\frac{\partial V_0}{\partial \textbf{\textit{x}} } -
\epsilon e^{st} 
\frac{\partial \tilde V}{\partial \textbf{\textit{x}} },
\label{eq:equation-motion-for-x-and-v}
\end{equation}
where $V_0(\textbf{\textit{x}})$ is the equilibrium potential 
field generated by galactic stars and a possible dark matter 
halo. In writing equation (\ref{eq:equation-motion-for-x-and-v}), 
I have assumed that the motion of stars is restricted to the disk 
plane. When the equilibrium state is axisymmetric and the dark 
component is spherical, $V_0$ becomes a function of radial distance 
to the galactic center and the unperturbed equations (with $\epsilon=0$)
are integrable. In such a circumstance, the phase space is filled 
by rosette orbits denoted by $[\textbf{\textit{x}}_0(t),\textbf{\textit{v}}_0(t)]$. 
The growth of perturbations, whatever the magnitude of $\epsilon e^{st}\ll 1$ 
may be, deforms stellar orbits. Orbital deformations are measured by 
$\tilde \textbf{\textit{x}}=\textbf{\textit{x}}-\textbf{\textit{x}}_0$ 
and $\tilde \textbf{\textit{v}}=\textbf{\textit{v}}-\textbf{\textit{v}}_0$, 
which can be used in (\ref{eq:equation-motion-for-x-and-v}) to obtain 
\begin{equation}
\frac{d\tilde \textbf{\textit{x}} }{dt}=\tilde \textbf{\textit{v}},~~
\frac{d\tilde \textbf{\textit{v}} }{dt}=- \left [
\frac{\partial^2 V_0}{\partial \textbf{\textit{x}}^2 }
\right ]_{ \textbf{\textit{x}}_0 } \!\!\!\! \cdot \tilde \textbf{\textit{x}} 
- \epsilon e^{st} \left [ 
\frac{\partial \tilde V}{\partial \textbf{\textit{x}} }
\right ]_{ \textbf{\textit{x}}_0 }.
\label{eq:equation-motion-for-tilde-x-and-v}
\end{equation}

Although a proper equilibrium distribution function $f_0(\textbf{\textit{x}},\textbf{\textit{v}})$ can self-consistently 
reproduce $\Sigma_0(\textbf{\textit{x}})$ using rosette orbits, 
the perturbed density $\Sigma_1(\textbf{\textit{x}},t)$ 
(corresponding to $V_1$) cannot be supported by rosette 
orbits alone and orbital deformations are necessary for 
the self-consistency of density waves.
According to equations (\ref{eq:equation-motion-for-x-and-v}) 
and (\ref{eq:equation-motion-for-tilde-x-and-v}), orbital 
deformations of ${\cal O}\left ( \epsilon e^{st} \right )$
are sufficient to support the growth of density/potential 
perturbations up to the same order of magnitude of such 
deformations over a time scale of $1/{\cal O}\left ( \epsilon e^{st} \right )$. 
As the time is elapsed, the amplitude of perturbations increases 
exponentially and the solution of the linearized collisionless 
Boltzmann equation (CBE) fails when $\epsilon e^{st}\sim 1$. 
During my mode calculations, I realized that the orbital axes 
of certain stars librate in a coordinate frame that rotates with 
the density pattern. This {\it resonant capture} initially seemed 
to be a higher-order nonlinear effect but further experiments 
showed that the resonant gap is constrained by the magnitude 
of density perturbations. The complex behavior of stars for 
infinitesimally small yet non-zero $\epsilon e^{st}\ll 1$, 
and the role of resonant stars in the generation of discrete 
galactic modes, are investigated in this paper. 

I use the results of Paper I and introduce a new dynamical
mechanism that sparks unstable modes and governs the singular
oscillations of \citet{vK55} modes. Resolving the origin of
instabilities and amplitude saturation precede my nonlinear
calculations, which were made feasible in Paper I by the
Petrov-Galerkin method and reducing the CBE to a system of 
nonlinear ordinary differential equations. Those reduced equations, 
however, are valid only when orbits are regular and averaging 
over angle variables is allowed. As unstable modes grow, chaotic 
orbits come into existence and the weighted residual form of the 
CBE must be modified to handle them. I quote some of the results 
of such modifications in this paper when I discuss the issue of 
mode saturation. In Paper III, I will give a full account of the 
mathematical and numerical modeling of stochastic layers, and will 
analyze modal interactions after their saturation phase.

For the cored exponential disk embedded in the field of the cored
logarithmic potential, I describe the decomposed Fourier components
of unstable bar and spiral modes in \S\ref{sec:mode-decomposition}
and highlight the existence of a phase shift between different
components. In \S\ref{sec:capture-into-resonances}, I derive a
condition for the corotation of the orbital axes of an ensemble of
stars and explain the role of such a synchronous motion in pattern
formation. I dedicate \S\ref{sec:perturbed-stellar-dynamics} to
exploring the orbital structure of a perturbed stellar disk and
identify a resonance mechanism that can generate both stationary and
growing modes. I reveal the mechanism of angular momentum transfer
between Fourier components and derive analytical expressions for 
the growth of resonance zone. I address the origin of instabilities 
in \S\ref{sec:origin-of-instabilities}, present a saturation
mechanism for unstable modes in \S\ref{sec:mode-saturation}, 
and discuss about the global stability of soft-centered and 
scale-free disks. I explain the restrictions of LBK's mode mechanism
in \S\ref{sec:discussion-and-conclusions} and end up the paper 
with concluding remarks.

\section{THE MODEL}
\label{sec:used-model}

The calculations of the present study are carried out for the cored
exponential disks of JH whose eigenfrequency spectra and mode shapes
have been completely explored in Paper I. The model has a dark
matter halo and the motion of stars in the equilibrium state is
governed by the cored logarithmic potential
\begin{equation}
V_0(R)=v_0^2 \ln \sqrt{1+R^2/R_C^2 },
\end{equation}
which is the resultant gravitational potential of luminous and
dark components. Throughout the paper, all length and velocity
variables are normalized, respectively, to the core radius of the
potential ($R_C$) and the asymptotic velocity of stars on circular
orbits ($v_0$) by setting $R_C=v_0=1$. To describe the physical
quantities in the configuration space, I will use the usual polar
coordinates $(R,\phi)$ and their Cartesian counterparts $(x,y)$
where $R$ is the radial distance from the galactic center and $\phi$
is the azimuthal angle. The specific model that I
adopt here is a relatively cold, near-maximal disk with no dark
matter concentration in the region with rising rotation curve. The
model parameters are set to $\left (N,\lambda,\alpha \right )=
(6,1,0.42)$ where $N$ is an integer exponent that controls 
the proportion of circular orbits and the disk temperature. Larger 
values of $N$ give rise to colder disks. For $N=6$, the parameter 
$Q$ of Toomre (1964) is marginally larger than 1. The parameter 
$\lambda$ is defined as the ratio $R_C/R_D$ with $R_D$ being the 
core radius of the equilibrium density. $\alpha$ is a factor 
that controls the total mass of the stellar component. The chosen 
model with $\alpha=0.42$ is near-maximal within a radius 
of $\approx 2.5 R_C$ and dominated by dark matter beyond it. 
The calculations of Paper I revealed the eigenspectrum of this 
model that includes a compact bar mode B1 and a sequence of 
spiral modes S1, $\cdots$, S6. I select modes B1 and S2 as my 
case studies of sections \ref{sec:mode-decomposition} through \ref{sec:perturbed-stellar-dynamics}. The reason for 
choosing mode S2 is its extensive and prominent spiral structure 
and modest growth rate. In section \ref{sec:mode-saturation},
I will also display some results for modes S1, S3 and S6.

\section{DECOMPOSITION OF UNSTABLE MODES}
\label{sec:mode-decomposition}

One of the advantages of the method developed in Paper I is that the
density function of a mode can be readily decomposed to its constituent 
Fourier components in angle-action space. The phase shifts between
density components determine the magnitude and direction of the
torque that is exerted on each component. So we can probe the
transfer of angular momentum and identify the direction of its flow
once a global mode develops.

The perturbed distribution and Hamiltonian functions of an unstable
mode can be expanded as (Paper I)
\begin{eqnarray}
f_1(\Theta,\textbf{\textit{J}},t) \! &=& \! {\rm Re} \!\!\!\!\!\!
\sum_{m,l=-\infty}^{\infty}\sum_{j=0}^{\infty} \! \epsilon
d^{ml}_j(t) \Phi^{ml}_j(\textbf{\textit{J}})e^{\imath
\left(m\theta_\phi+l\theta_R \right)},
\label{eq:expansion-f1} \\
{\cal H}_1(\Theta,\textbf{\textit{J}},t) \! &=& \! {\rm Re} \!\!\!\!\!\!
\sum_{m,l=-\infty}^{\infty}\sum_{j=0}^{\infty} \! \epsilon
b^{ml}_j(t) \Psi^{ml}_j(\textbf{\textit{J}})e^{\imath
\left(m\theta_\phi+l\theta_R \right)}, \label{eq:expansion-V1}
\end{eqnarray}
where $\Theta=\left (\theta_R,\theta_{\phi} \right )$ and
$\textbf{\textit{J}}=\left (J_R,J_{\phi} \right )$ are the angle
and action variables, respectively. The angles are defined based
on the radial and azimuthal frequencies
\begin{equation}
{\bf \Omega}=\left (\Omega_R,\Omega_{\phi} \right )=
\left ( \frac{\partial {\cal H}_0}{\partial J_R},
        \frac{\partial {\cal H}_0}{\partial J_{\phi}}
\right ),
\end{equation}
of stars on rosette orbits so that
\begin{equation}
\dot \theta_R = \Omega_R(\textbf{\textit{J}}),~~
\dot \theta_{\phi} = \Omega_{\phi}(\textbf{\textit{J}}).
\end{equation}
A dot stands for the time derivative and
${\cal H}_0(\textbf{\textit{J}})$ is the integrable Hamiltonian
of the axisymmetric equilibrium state. For a normal mode the
amplitude functions $d^{ml}_j(t)$ and $b^{ml}_j(t)$ depend on
the time variable $t$ through the simple exponential law
$\exp(-{\imath}\omega t)$ with $\omega=m\Omega_p+{\imath}s$ and
$\imath=\sqrt{-1}$. Here $\Omega_p$ and $s$ are the pattern speed
and growth rate of an unstable mode of angular wavenumber $m$.
On the other hand, the perturbed potential function $V_1$ and
its associated surface density $\Sigma_1$ can be expanded in
the configuration space as
\begin{eqnarray}
V_1(R,\phi,t) &=& {\rm Re} \!\!\!\!\!
\sum_{m=-\infty}^{\infty}\sum_{j=0}^{\infty} \epsilon
a^{m}_j(t) \psi^{\vert m\vert}_j(R) e^{\imath m\phi},
\label{eq:expansion-V1-configuration} \\
\Sigma_1(R,\phi,t) &=& {\rm Re} \!\!\!\!\!
\sum_{m=-\infty}^{\infty}\sum_{j=0}^{\infty} \epsilon
a^{m}_j(t) \sigma^{\vert m\vert}_j(R) e^{\imath m\phi},
\label{eq:expansion-Sigma1-configuration}
\end{eqnarray}
where $\sigma^{\vert m\vert}_j(R)$ and $\psi^{\vert m\vert}_j(R)$
are surface density and potential basis functions, respectively.
They satisfy Poisson's integral and the bi-orthogonality
condition
\begin{equation}
D_j(m) \delta_{j,j'}\delta_{m,m'}=
2\pi \int\limits_{0}^{\infty} \psi^{\vert m\vert}_j(R)
\sigma^{\vert m'\vert}_{j'}(R) RdR.
\end{equation}
Here $\delta_{m,m'}$ is the Kronecker delta and $D_j(m)$
are some constants that depend on our special choice of
basis functions. Noting $V_1\equiv {\cal H}_1$, one can
equate (\ref{eq:expansion-V1}) and
(\ref{eq:expansion-V1-configuration}), multiply the identity
by $\exp[-\imath(l\theta_R+m\theta_{\phi})]$ and integrate
over the angle variables to obtain
\begin{equation}
\sum_{j=0}^{\infty} b^{ml}_j(t)
\Psi^{ml}_j(\textbf{\textit{J}})=
\sum_{j=0}^{\infty} a^{m}_j(t)
\tilde \Psi^{ml}_j(\textbf{\textit{J}}),
\end{equation}
\begin{equation}
\tilde \Psi^{ml}_j(\textbf{\textit{J}})\!=\!
\frac{1}{\pi}\int\limits_{0}^{\pi} \! \psi^{\vert m\vert}_j(R)\cos
\left [ l\theta_R+m\left (\theta_{\phi}-\phi \right ) \right ]
d\theta_R.
\end{equation}
Setting $\Psi^{ml}_j(\textbf{\textit{J}})=
\tilde \Psi^{ml}_j(\textbf{\textit{J}})$
gives $b^{ml}_j(t)=a^m_j(t)$, which is a remarkable
simplification. A relation between $a^m_j(t)$ and
$d^{ml}_j(t)$ then follows from the weighted residual
form of the fundamental equation
\begin{equation}
f_1(\Theta,\textbf{\textit{J}},t)d\textbf{\textit{J}}
d\Theta=\Sigma_1(R,\phi,t)RdRd\phi.
\end{equation}
I obtain
\begin{eqnarray}
a^m_j(t)&=& \frac
{4\pi^2}{D_j(m)}\sum_{l=-\infty}^{\infty}\sum_{k=0}^{\infty}
\Lambda^{ml}_{jk}d^{ml}_k(t),\label{eq:a-versus-d} \\
\Lambda^{ml}_{jk} &=& \int\limits \Psi^{ml}_j(\textbf{\textit{J}})
\Phi^{ml}_k(\textbf{\textit{J}}) d \textbf{\textit{J}},
\label{eq:define-Lambda}
\end{eqnarray}
where
\begin{equation}
\Phi^{ml}_k(\textbf{\textit{J}})=
     \frac{l\left (\partial f_0/\partial J_R\right )
      +m\left (\partial f_0/\partial J_{\phi} \right )}
     {l\Omega_R+m\Omega_{\phi}} 
     \Psi^{ml}_k(\textbf{\textit{J}}).
     \label{eq:define-Phi-vs-Psi}
\end{equation}

Let ${\cal L}$ be the total angular momentum of the disk, which must
be conserved in an isolated galaxy. This means that the torque
$d{\cal L}/dt$ must vanish. Defining
\begin{eqnarray}
a^m_j(t) &=& e^{-{\imath}\omega t} \tilde a^m_j=e^{-{\imath}\omega t}
\left ( u^m_j+{\imath}v^m_j \right ), \\
d^{ml}_j(t) &=& e^{-{\imath}\omega t} \tilde d^{ml}_j =
e^{-\imath \omega t} \left ( U^{ml}_j+{\imath}V^{ml}_j \right ),
\end{eqnarray}
one obtains
\begin{eqnarray}
{d{\cal L}\over dt} &=& 2m \pi^2 \epsilon ^2 
e^{2st}\sum_{l=-\infty}^{\infty}
L_m(l)=0, \label{eq:total-torque}\\
L_m(l) &=& \sum_{j,k=0}^{\infty} \Lambda^{ml}_{jk}\left (v^m_j
U^{ml}_k -u^m_j V^{ml}_k \right ). \label{eq:torque-component-1}
\end{eqnarray}
The constant vectors $\left (U^{ml}_k,V^{ml}_k \right )$ and $\left
(u^m_j,v^m_j \right )$ are obtained from the linear eigenvalue
equations of Paper I and a subsequent use of equation
(\ref{eq:a-versus-d}). Equation (\ref{eq:total-torque}) is analogous
to equation (B6) in JH and $L_m(l)$ shows the rate of angular
momentum flow into/from the $l$th Fourier component (in the
radial angle $\theta_R$) whose corresponding density in the
configuration space is
\begin{equation}
\Sigma^{l}_1=\! \epsilon e^{st} {\rm Re} \!\!
\sum_{j,k=0}^{\infty} \!  \frac{4\pi^2}{D_j(m)}
\Lambda^{ml}_{jk}\sigma^{|m|}_j(R) \tilde d^{ml}_k
e^{{\imath}m\left ( \phi-\Omega_p t \right )}.
\end{equation}
The angular momentum conservation of the disk implies
that some $L_m(l)$ take negative and some other positive values with
a total vanishing sum. This means that $\Sigma^{l}_1$, the pattern
corresponding to $L_m(l)$, is subject to a positive torque from
other components if $L_m(l)>0$, and a negative torque otherwise.

\begin{figure*}
\centerline{\hbox{\includegraphics[width=0.9\textwidth]
{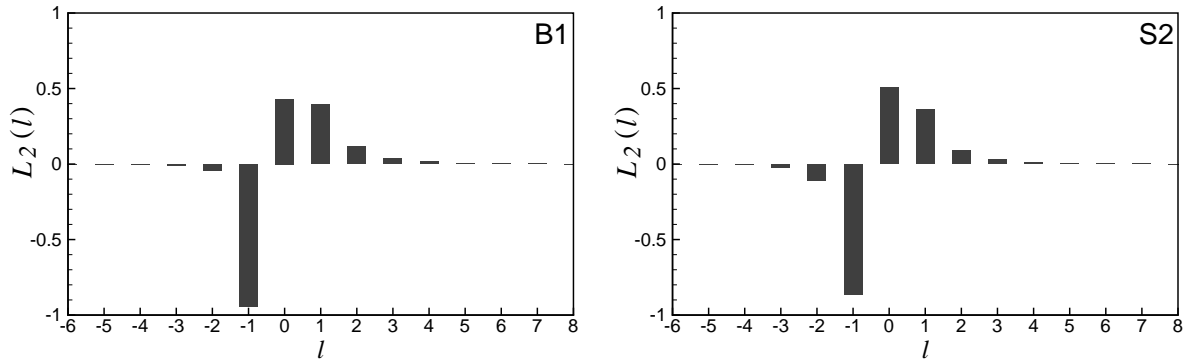} }}
\caption{The angular momentum content of different
Fourier components of modes B1 and S2. The sum
$\sum_{l=-\infty}^{+\infty}L_m(l)$ vanishes because
the total angular momentum of the disk is conserved.
The function $L_m(l)$ is normalized so that the sum of
positive components is unity.
\label{pic:ang-momentum-modes-B1-S2} }
\end{figure*}
\begin{figure*}
\centerline{\hbox{\includegraphics[width=0.9\textwidth]
{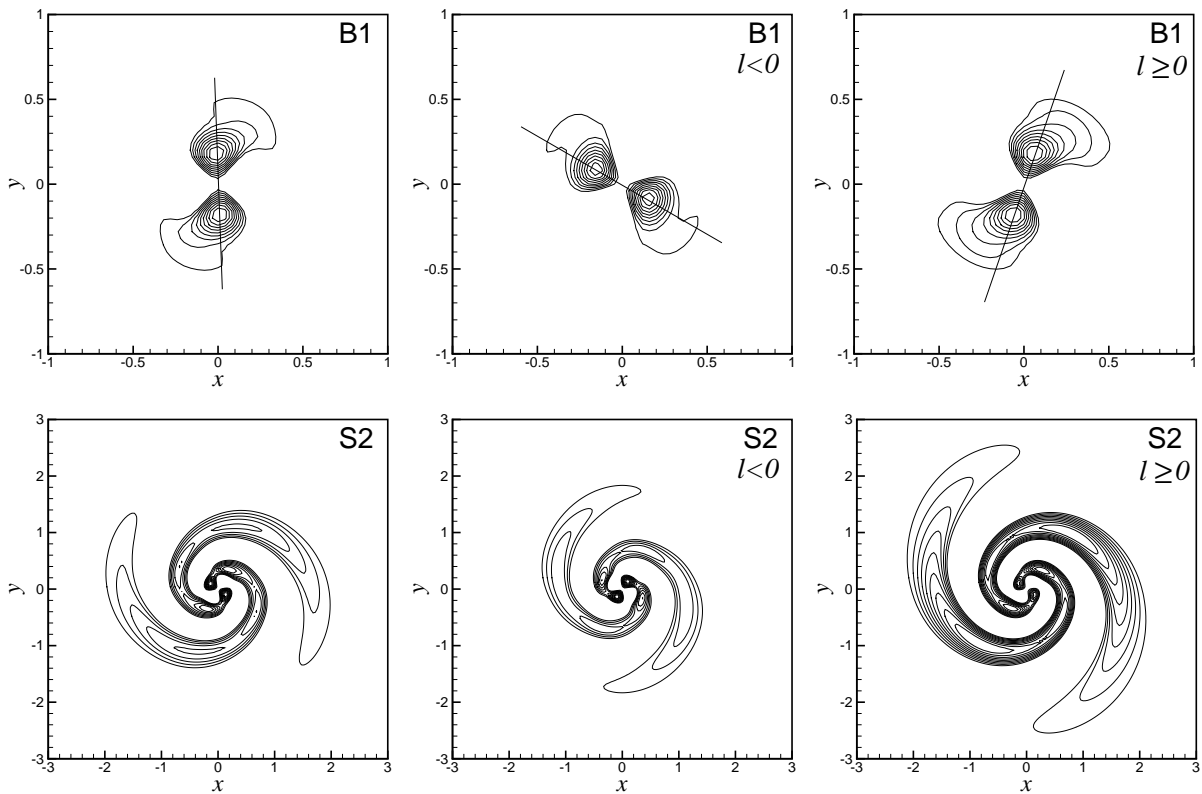} }} \caption{Modes B1 and S2 and their
components that release and absorb angular momentum. Left panels
show the mode shapes, which rotate counter-clockwise. The pattern
speeds and growth rates of modes B1 and S2 are
$(\Omega_p,s)=(0.918,1.160)$ and $(\Omega_p,s)=(0.454,0.216)$,
respectively. Middle and right panels demonstrate the
$\Sigma^{l-}_1$ and $\Sigma^{l+}_1$ componenets, respectively. Solid
lines in top panels highlight the orientations of the bar mode and
its components. The prominent phase shift between $\Sigma^{l-}_1$
and $\Sigma^{l+}_1$ is responsible for the gravitational torque
between them. The isocontours show the positive part of the density
from 10\% to 90\% of the maximum with increments of 10\%.
\label{pic:components-modes-B1-S2}}
\end{figure*}

Figure \ref{pic:ang-momentum-modes-B1-S2} shows the variation
of $L_2(l)$ versus $l$ for modes B1 and S2. The results
are in accordance with the bar charts of JH: components with $l<0$
lose angular momentum and those with $l\ge 0$ gain it.
As JH had already pointed out, a few components ensure the
convergence of Fourier series in the $\theta_R$-direction.
The superposition of the $l\ge 0$ components of the perturbed
density, defined as
\begin{equation}
\Sigma^{l+}_1(R,\phi,t)=\sum_{l=0}^{\infty}\Sigma^{l}_1(R,\phi,t),
\end{equation}
will thus be a pattern that experiences a positive torque exerted by
\begin{equation}
\Sigma^{l-}_1(R,\phi,t)=\sum_{l=-\infty}^{-1}\Sigma^{l}_1(R,\phi,t).
\end{equation}
In response, the reaction torque of $\Sigma^{l+}_1$ will drain the
angular momentum of $\Sigma^{l-}_1$ so that ${\cal L}$ is conserved.
Figure \ref{pic:components-modes-B1-S2} shows modes B1 and S2 and
their components that emit and absorb angular momentum. It is seen
that in both modes, the phase of $\Sigma^{l+}_1$ lags that of
$\Sigma^{l-}_1$ by a magnitude of $\approx 90^{\circ}$ and the
positive parts of $\Sigma^{l-}_1$ fill the regions of negative
$\Sigma^{l+}_1$. Due to this phase shift, the angular momentum is
transferred between $l<0$ and $l\ge 0$ components and a
counterclockwise torque is exerted on $\Sigma^{l+}_1$. This
phenomenon is more obvious in mode B1 that has a definite spatial
orientation. The most important question relevant to the origin of
instabilities arises now: Are the density components
$\Sigma^{l}_1(R,\phi,t)$ generated by stellar orbits at
resonances? Analytical calculations of sections
\ref{sec:capture-into-resonances} and
\ref{sec:perturbed-stellar-dynamics} provide the answer.

\section{SYNCHRONOUS PRECESSION OF THE ORBITAL AXES}
\label{sec:capture-into-resonances}

The perturbed density $\Sigma_1$ of a mode with the pattern
speed $\Omega_p$ must be supported by the slow motion of stars
in a rotating coordinate frame of angular velocity $\Omega_p$.
In other words, by sitting on a moving coordinate system one
must be able to identify a group of stars that are losing
their angular velocity, and another group of the same kind
but with different initial conditions, gaining it. A density
peak is expected in the region that the angular velocity of
the slow ensemble is minimum. Since the angular velocity of
a star is minimum at its orbital apocenter, the precession
rate of the orbital axes of participating stars in the 
pattern formation should be in resonance with the pattern
speed. The above scenario is legitimate as long as pattern
stars stay far from Lindblad resonances that can generate  
higher density regions due to the radial slowing of stars. 
My computations show that all unstable modes fulfill this 
requirement and azimuthal slowing is the main origin of 
density perturbations (see \S\ref{sec:perturbed-stellar-dynamics}).

The evolution of $\phi(t)$ is crucial for understanding the slow
dynamics in the configuration space, but we usually have the perturbed
distribution and potential functions in the angle-action space. It is
thus useful to find how $\phi$ depends on $(\Theta,\textbf{\textit{J}})$.
The simplest relation can be obtained by
expanding $\exp[\imath m (\phi-\theta_{\phi})]$ in Fourier series
of $\theta_R$ as
\begin{equation}
e^{ {\imath} m \left (\phi-\theta_{\phi}\right )} =
\sum_{l=-\infty}^{+\infty}
\xi_{ml}\left ( \textbf{\textit{J}} \right )e^{{\imath}l
\theta_R}, \label{eq:relation-between-phi-angles}
\end{equation}
where
\begin{equation}
\xi_{ml}(\textbf{\textit{J}})=
\frac 1{2\pi}\oint \cos \left [ l\theta_R+
m\left (\theta_{\phi}-\phi \right )\right ] d\theta_R.
\label{eq:fourier-coeff-phi-vs-angles}
\end{equation}
The functions $\xi_{ml}(\textbf{\textit{J}})$ are computed by 
integrating along rosette orbits (over a period of radial 
oscillation), and they vanish for circular orbits when $l\not=0$ 
and for radial orbits when $m\not =-2l$. 
Multiplying equation (\ref{eq:relation-between-phi-angles})
by $\exp[-\imath m(\phi-\theta_{\phi})]$ and integrating the
identity over a period of $\theta_R$ yields the
following useful relation
\begin{equation}
\sum_{l=-\infty}^{+\infty}
\left [\xi_{ml}(\textbf{\textit{J}})\right ]^2=1.
\label{eq:square-fourier-identity}
\end{equation}
Moreover, taking the partial derivative of
(\ref{eq:relation-between-phi-angles}) with respect
to $\theta_R$, multiplying both sides of the resulting
equation by $\exp[-\imath m(\phi-\theta_{\phi})]$ followed
by an integration over $\theta_R$, leads to (Scott Tremaine,
private communication)
\begin{equation}
\sum_{l=-\infty}^{+\infty}
l \left [\xi_{ml}(\textbf{\textit{J}})\right ]^2=0.
\label{eq:l-multiply-square-fourier-identity}
\end{equation}

I now multiply (\ref{eq:relation-between-phi-angles}) by
$\exp(\imath m \theta_{\phi})$, differentiate both sides
of the resulting equation with respect to $t$ and obtain
\begin{eqnarray}
m\dot \phi &=& -{\rm Re} \!\! \sum_{l=-\infty}^{+\infty} \!\!
\imath \left ( \frac {\partial \xi_{ml} }{\partial J_R}
        \dot J_R+
    \frac {\partial \xi_{ml} }{\partial J_{\phi} }
        \dot J_{\phi}
\right ) e^{\imath \left (l\theta_R+m\theta_{\phi}-m\phi \right )}
\nonumber \\
&+&
{\rm Re} \sum_{l=-\infty}^{+\infty} \!\! \xi_{ml}
\left ( l\dot \theta_R +m\dot \theta_{\phi} \right )
e^{\imath \left (l\theta_R+m\theta_{\phi}-m\phi \right )}.
\label{eq:diff-relation-phi-angles}
\end{eqnarray}
The perturbed motions of stars are governed by Hamilton's
equations
\begin{equation}
\dot \Theta = {\bf \Omega}(\textbf{\textit{J}}) +
\frac{\partial {\cal H}_1}{\partial \textbf{\textit{J}}},~~
\dot \textbf{\textit{J}} =
- \frac{\partial {\cal H}_1}{\partial \Theta},
\end{equation}
that can be used to rewrite (\ref{eq:diff-relation-phi-angles})
in the form
\begin{eqnarray}
m\dot \phi &=& {\rm Re}
\sum_{l=-\infty}^{+\infty} \!\!
\xi_{ml} \left [
l\Omega_R+m\Omega_{\phi} \right ]
    e^{\imath \left (
           l\theta_R+m\theta_{\phi}-m\phi
              \right ) } \nonumber \\
&-& {\rm Re} \sum_{l=-\infty}^{+\infty}
\imath e^{-\imath m\phi}
\left [ e^{{\imath}(l\theta_R+m\theta_{\phi} )}
\xi_{ml},{\cal H}_1  \right ],
\label{eq:relation-diff-phi-angles}
\end{eqnarray}
where $[\cdots,\cdots]$ denotes a Poisson bracket taken over the
angle-action space. On substituting for the complex conjugate of
$\exp(\imath m\phi)$ from (\ref{eq:relation-between-phi-angles}) in
(\ref{eq:relation-diff-phi-angles}) and noting that the second term
on the right hand side of (\ref{eq:relation-diff-phi-angles}) is of
${\cal O}\left (\epsilon e^{st} \right )$, one finds
\begin{equation}
m\dot \phi = \! {\rm Re} \!\!\!
\sum_{l,k=-\infty}^{+\infty} \!\!
\!\! \xi_{ml} \xi_{mk} \left [
l\Omega_R+m\Omega_{\phi} \right ]
    e^{\imath (l-k)\theta_R }+{\cal O}
    \left (\epsilon e^{st}\right ),
\label{eq:relation-diff-phi-theta-r}
\end{equation}
which shows the functional dependence of $\dot \phi$ on the
actions and the radial angle.

Equation (\ref{eq:relation-diff-phi-theta-r}) can be
written as
\begin{equation}
m\dot \phi = m\Omega_{\phi}(\textbf{\textit{J}}) +
\eta(\textbf{\textit{J}},\theta_R)+
{\cal O}\left ( \epsilon e^{st} \right ),
\label{eq:simple-relation-for-dot-phi}
\end{equation}
where the constant part $m\Omega_{\phi}$ is obtained by
setting $l=k$ in the double summation of
(\ref{eq:relation-diff-phi-theta-r}) and subsequent
application of relations (\ref{eq:square-fourier-identity})
and (\ref{eq:l-multiply-square-fourier-identity}).
$\eta(\textbf{\textit{J}},\theta_R)$ is a periodic
function of $\theta_R$ that stands for all other terms
of (\ref{eq:relation-diff-phi-theta-r}) with $l\not =k$.
$\Omega_{\phi}(\textbf{\textit{J}})$ is thus the precession
rate of the orbital axis of stars whose energy and angular
momentum correspond to the action vector $\textbf{\textit{J}}$.
Those stars can contribute to a developing density peak when 
they linger at the apocenter of their orbits. 
Such a mass deposition will continue in a rotating frame 
of angular velocity $\Omega_p$ and over the time scale 
$1/{\cal O}\left (\epsilon e^{st} \right )$ if the condition 
for synchronous 
precession
\begin{equation}
\left \langle \dot\phi \right \rangle _{\theta_R} 
\!\!\! -\Omega_p \approx {\cal O} \left (\epsilon e^{st} \right ), 
\label{eq:precession-frame-resoannce}
\end{equation}
holds. By defining
\begin{equation}
\mu^l_m(\textbf{\textit{J}})=
l\Omega_R(\textbf{\textit{J}})+
m\Omega_{\phi}(\textbf{\textit{J}})-m\Omega_p,
\label{eq:define-mu}
\end{equation}
and using (\ref{eq:simple-relation-for-dot-phi}), the
condition (\ref{eq:precession-frame-resoannce})
takes the convenient form
\begin{equation}
\mu^0_m(\textbf{\textit{J}})
\approx {\cal O} \left ( \epsilon e^{st} \right ),
\label{eq:condition-secular-resonance}
\end{equation}
which is the definition of the CR. The ILR and OLR occur
if $\vert \mu^l_m\vert$ diminishes for $l=-1$ and $l=1$,
respectively.

\begin{figure}
\centerline{\hbox{\includegraphics[width=0.45\textwidth]
{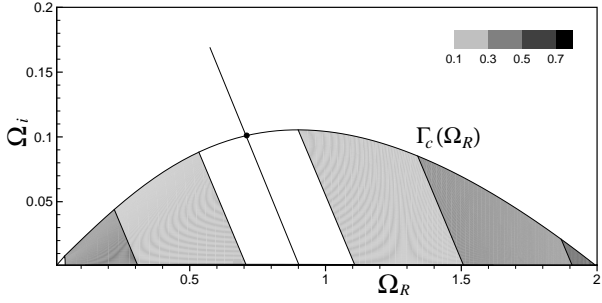} }} \caption{The frequency space of the cored
logarithmic potential with $R_C=v_0=\Omega_0=1$. The upper boundary
$\Omega_i=\Gamma_c(\Omega_R)$ corresponds to circular orbits with
$J_R=0$. Contours show the variation of the CR indicator $\vert
\mu^0_2({\bf \Omega})\vert$ over the frequency space for
$\Omega_p=0.454$. Lighter regions correspond to smaller values of
$\vert \mu^0_2({\bf \Omega})\vert$. The function $\mu^0_2({\bf
\Omega})=2(\Omega_{\phi}-\Omega_p)$ vanishes at the intersection of
the drawn straight line and the frequency space. The star at the
intersection of the line $\mu^0_2({\bf \Omega})=0$ and the upper
boundary moves on the corotation circle. \label{pic:CR-zones}}
\end{figure}

A graphical representation of (\ref{eq:condition-secular-resonance})
will help to sharpen our understanding of the ensemble of stars that
can support a rotating pattern. Since there is a one-to-one, onto 
and invertible map $\textbf{\textit{J}} \rightarrow {\bf \Omega}$
for initially axisymmetric disks (excluding Keplerian and harmonic
oscillator potentials), any functional form of the actions can be
described in terms of ${\bf \Omega}$ as well. For instance, one may
write $\mu^0_{m}(\textbf{\textit{J}})\equiv \mu^0_{m}({\bf \Omega})$.
I therefore identify resonant regions in the frequency space.
For demonstrating the distribution of dependent quantities,
I will use the pair $\left (\Omega_R,\Omega_i \right )$ as the
coordinates where $\Omega_i=\Omega_{\phi}-\frac 12 \Omega_R$.
The reason for this choice is that the orbital frequencies of
soft-centered stellar disks fill a very narrow region in the 
${\bf \Omega}$-space (Hunter 2002, hereafter H02; Figure 1 in JH), 
which does not provide enough resolution for the visual identification 
of some fine structures. 
Denoting $\Omega_0=\left [\Omega_{\phi}\right ]_{\rm max}$
in a cored stellar disk, the center of the disk and infinity
correspond to $(\Omega_{R},\Omega_i)=(2\Omega_0,0)$ and
$(\Omega_{R},\Omega_i)=(0,0)$, respectively. Figure \ref{pic:CR-zones}
shows the frequency space of the cored logarithmic potential.
The lower boundary ($\Omega_R$-axis) corresponds to radial orbits
with $J_{\phi}=0$ and the upper curved boundary
$\Omega_i=\Gamma_c(\Omega_R)$ is determined by the orbital
frequencies of circular orbits with $J_R=0$.

I have plotted in Figure \ref{pic:CR-zones} the countours of $\vert
\mu^0_{m}({\bf \Omega})\vert$ for $m=2$ and $\Omega_p=0.454\Omega_0$,
which is the pattern speed of mode S2. Lighter regions mark
the stars that are closer to the exact CR. The intersection of
the straight line $\mu^0_2({\bf \Omega})=0$ with the frequency
space corresponds to the stars at the CR. The centerline of the
CR zone begins at the location of a star moving on the corotation
circle, and extends into inner regions. 

By decreasing $\Omega_p$, the CR zone is pushed to the outskirts
of the disk where orbital frequencies are small and the surface density
has dropped substantially. Furthermore, the line $\mu^{0}_m({\bf \Omega})=0$
will not intersect the frequency space, and there will not be a CR if
$\Omega_p>\Omega_0$. For a similar reason, the ILR will be absent
if
\begin{eqnarray}
&{}& \Omega_p > \Omega_{\rm ILR}(m), \label{eq:condition-for-NO-ILR} \\
&{}& \Omega_{\rm ILR}(m) = \left [ \Gamma_c(\Omega_R) +\left (\frac m2 -1 \right )
\Omega_R \right ]_{\rm max}.
\label{eq:define-Omega-ILR}
\end{eqnarray}
Inequality (\ref{eq:condition-for-NO-ILR}) is satisfied 
when the line $\mu^{-1}_m({\bf \Omega})=0$ does not cross 
the boundary curve $\Omega_i=\Gamma_c(\Omega_R)$. 
These properties appear to correlate with the computed 
eigenfrequencies of unstable modes: according to the 
eigenspectra of Paper I, the pattern speeds of growing 
modes lie in the interval
\begin{equation}
\Omega_{\rm ILR}(m) < \Omega_p < \Omega_0.
\label{eq:constraint-pattern-speed-unstable}
\end{equation}
The lower limit may be violated by cold disks and disks with
inner cutouts (see \S\ref{sec:role-of-ILR}). The upper limit is 
violated by some rapidly rotating modes in galaxies with dark 
matter, and by mode B1 of cutout disks. After exploring the orbital 
structure of perturbed disks in \S\ref{sec:perturbed-stellar-dynamics}, 
I will explain the constraint (\ref{eq:constraint-pattern-speed-unstable}) 
and the exceptional cases that may not satisfy it.

\begin{figure*}
\centerline{ \hbox{\includegraphics[width=0.33\textwidth]{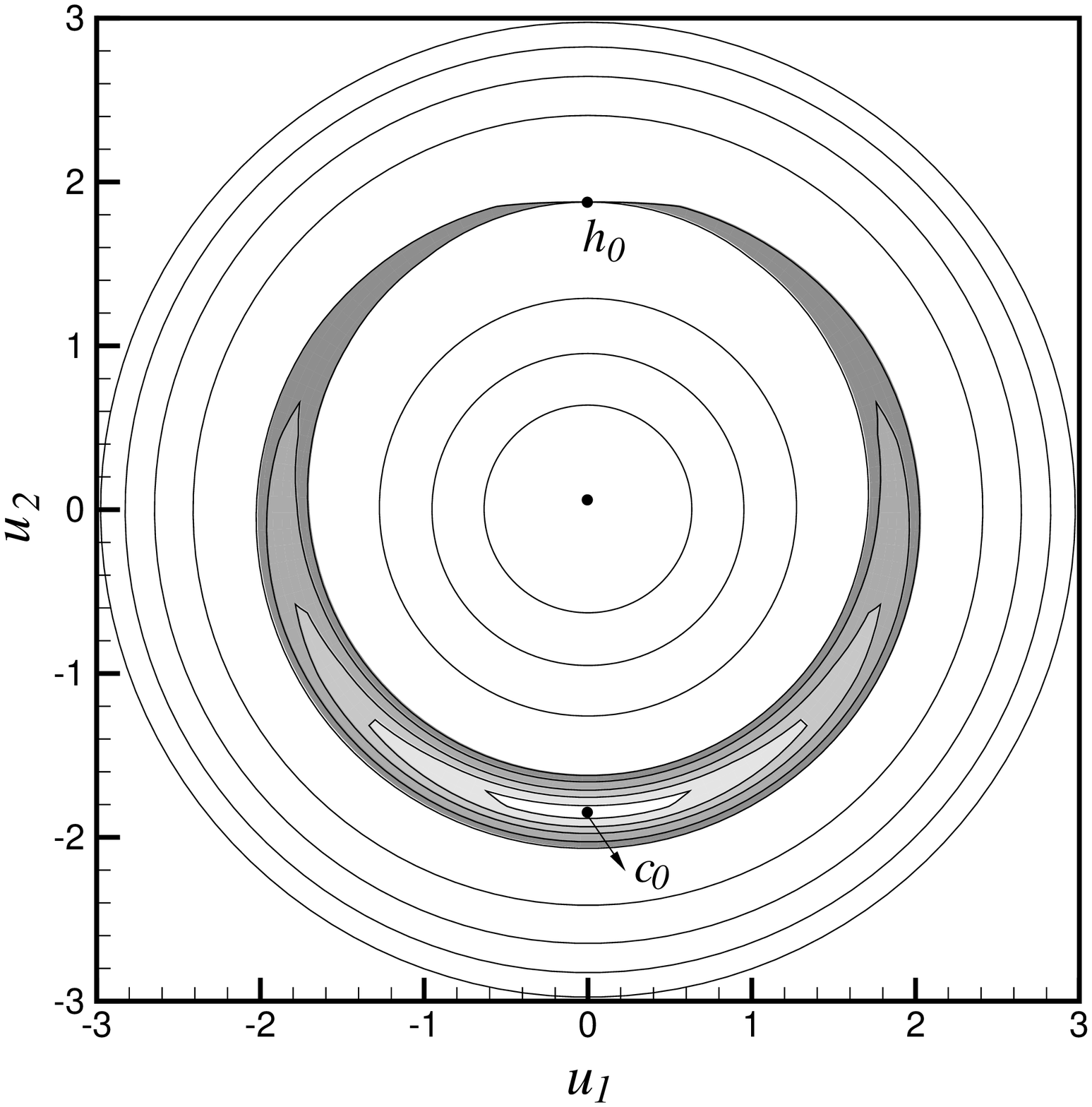}}
             \hbox{\includegraphics[width=0.33\textwidth]{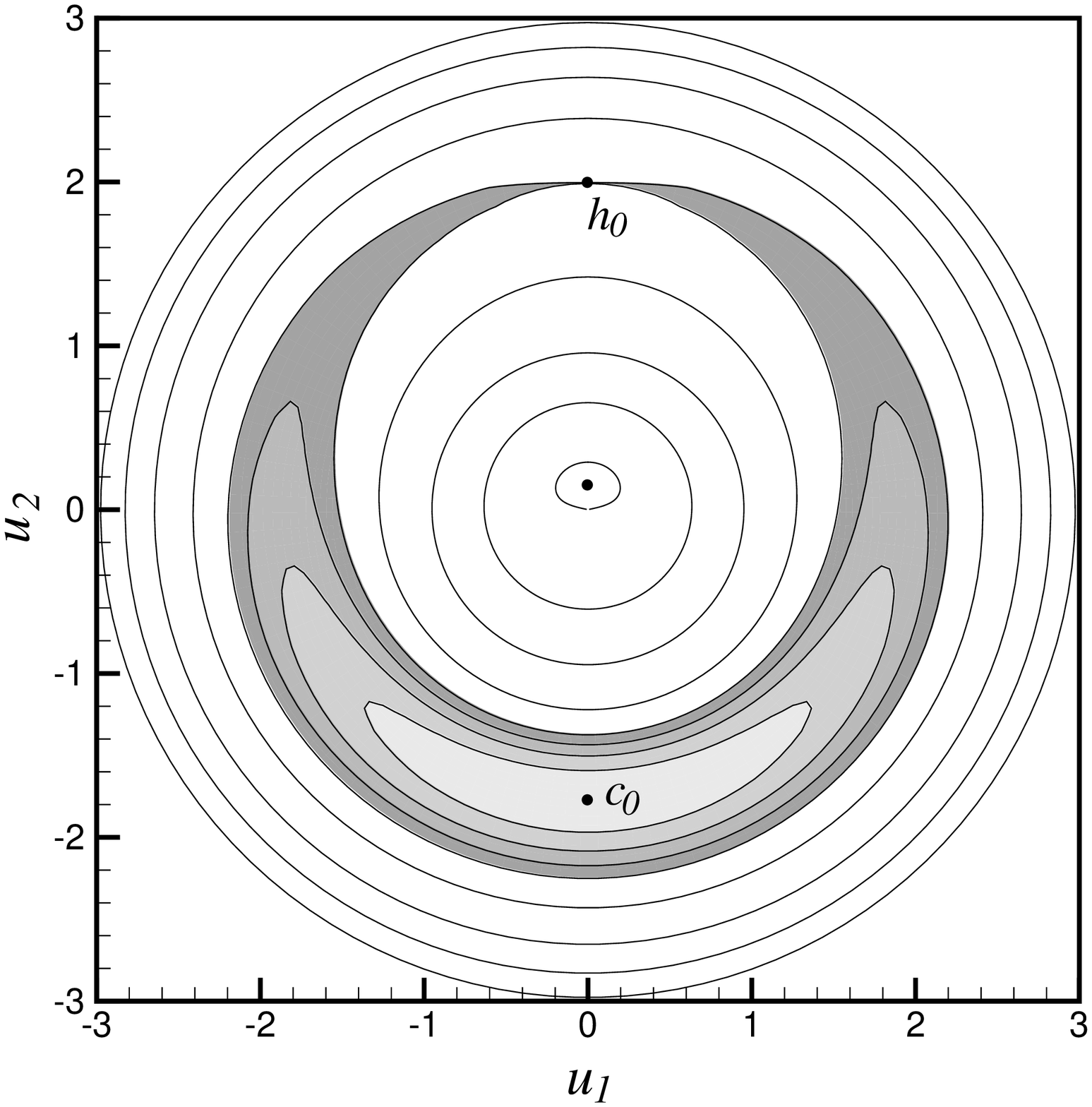}}
	     \hbox{\includegraphics[width=0.33\textwidth]{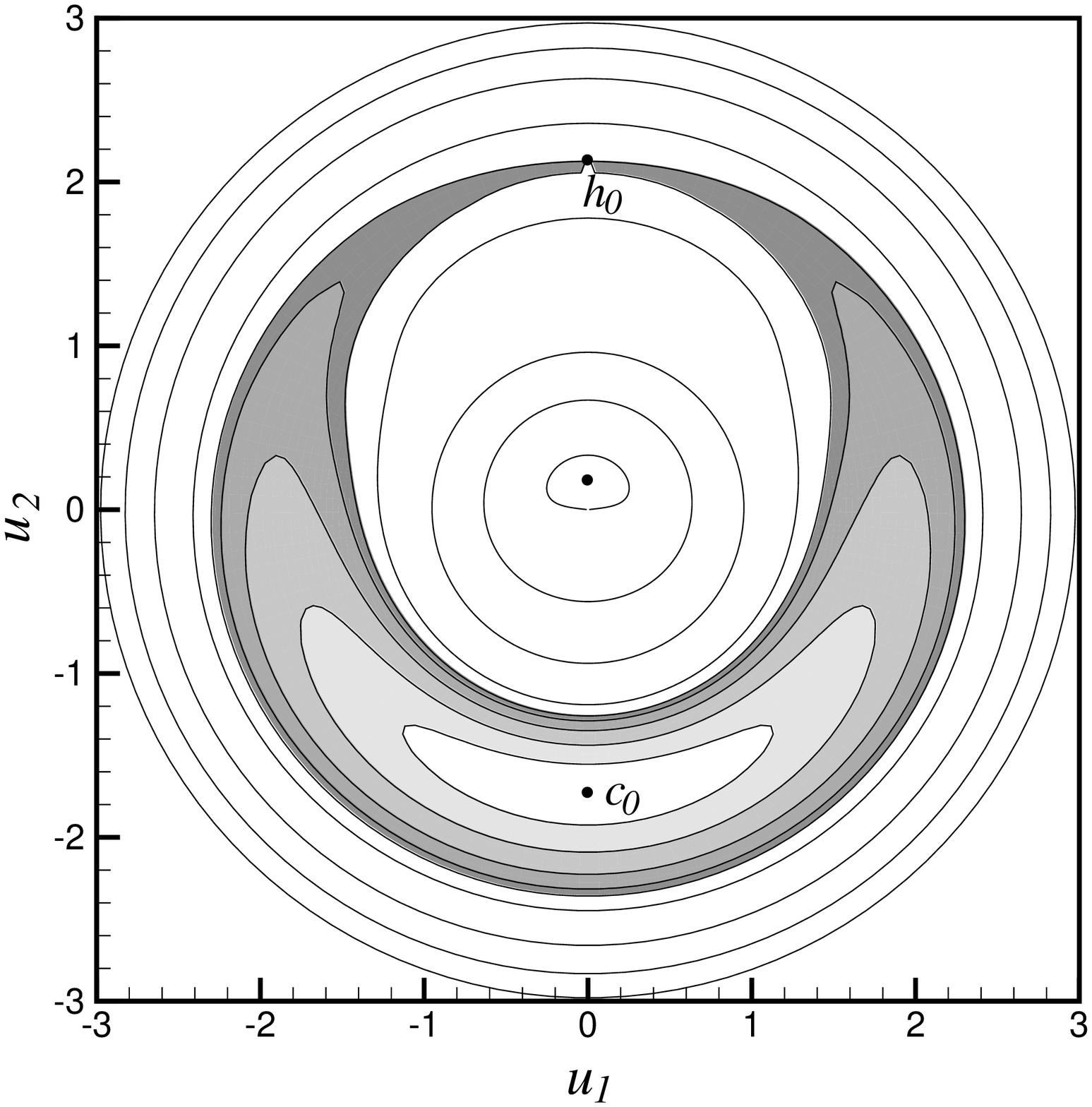}} } 
\caption{The instantaneous isocontours of the averaged Hamiltonian 
$\overline {\cal K}(2,0)$ for mode S2. The factor $\epsilon e^{st}$ 
is dealt with as a constant and has been set to 0.01 ({\em left panel}), 
0.05 ({\em middle panel}) and $0.1$ ({\em right panel}). 
The maximum of $\overline {\cal K}_1$ has been normalized 
to unity. The crescent-like resonant zone, driven by the 
instantaneous elliptic point $c_0$, has been shaded. Darker 
regions correspond to smaller values of $\overline {\cal K}(2,0)$. 
The separatrices intersect at the instantaneous hyperbolic 
point $h_0$. 
\label{pic:resonance-cavity}}
\end{figure*}

\section{STELLAR DYNAMICS IN THE PERTURBED DISK}
\label{sec:perturbed-stellar-dynamics}

The orbital dynamics of stars can be understood
through investigating the induced dynamics by each Fourier
component. We usually split the phase space to
several subspaces by using a Fourier expansion in $\theta_R$,
and each subspace is associated with a pattern component
(e.g., Figure \ref{pic:components-modes-B1-S2}). In the 
linear regime, the perturbed motion of stars due to the 
$l$th component can be traced by carrying out a canonical 
transformation 
\begin{equation}
\left (\theta_R,\theta_{\phi},J_R,J_{\phi} \right )
\rightarrow \left ( w_1,w_2,I_1,I_2 \right ),
\end{equation}
defined by the generating function \citep{L-B93}
\begin{equation}
{\cal S}=\left (l \theta_R+
     m\theta_{\phi}- m\Omega_p t \right )I_1+\theta_R I_2.
\end{equation}
This yields the transformation rules
\begin{equation}
\textbf{\textit{w}} = \frac{\partial {\cal S}}
                           {\partial \textbf{\textit{I}}},~~
\textbf{\textit{J}} =
\frac{\partial {\cal S}}{\partial \Theta},
\end{equation}
and the new Hamiltonian
\begin{equation}
{\cal K} = {\cal H}_0(\textbf{\textit{I}})+
{\cal H}_1(\textbf{\textit{w}},\textbf{\textit{I}},t)
+\partial {\cal S}/\partial t,
\end{equation}
so that
\begin{eqnarray}
{\cal K} &=& {\cal H}_0(\textbf{\textit{I}})-m \Omega_p I_1
\nonumber \\
 &+& \epsilon e^{st} ~{\rm Re} \! \sum_{k=-\infty}^{\infty} \sum_{j=0}^{\infty}
\tilde a^{m}_j \Psi^{mk}_j(\textbf{\textit{I}}) e^{{\imath} \left [
w_1 +\left (l-k \right )w_2 \right ]}.
\label{eq:new-K-Hamiltonian}
\end{eqnarray}
From the equations of motion
\begin{equation}
\dot \textbf{\textit{I}}=-\frac {\partial {\cal K} }
{\partial \textbf{\textit{w}} },~~
\dot \textbf{\textit{w}}=\frac {\partial {\cal K} }
{\partial \textbf{\textit{I}} },
\end{equation}
one can verify that $w_2=\Omega_R t+{\cal O}(\epsilon e^{st})$ 
is an increasing function of time, which can be made a cyclic 
coordinate by averaging ${\cal K}$ over $w_2$. Consequently, 
the action $I_2=J_R-l I_1=J_R-l J_{\phi}/m$ becomes an 
adiabatic invariant so that $\dot I_2={\cal O}(\epsilon^2 e^{2st})$, 
and the dynamics is reduced to the flows governed by the averaged 
Hamiltonian
\begin{eqnarray}
\overline {\cal K}(m,l) &\equiv& \langle {\cal K} \rangle _{w_2} =
\overline {\cal K}_0+\epsilon e^{st} \overline {\cal K}_1, 
\label{eq:averaged-bar-K} \\
\overline {\cal K}_0(\textbf{\textit{I}}) &=& 
{\cal H}_0(\textbf{\textit{I}})-m\Omega_p I_1 \\
\overline {\cal K}_1 &=& A^{l}_{m}(\textbf{\textit{I}})
\cos \left [ w_1+\vartheta^{l}_{m}(\textbf{\textit{I}}) \right ],
\label{eq:averaged-bar-K-w}
\end{eqnarray}
where I have kept the real (physical) part of
$\langle {\cal K} \rangle _{w_2}$ and
\begin{eqnarray}
A^{l}_{m}(\textbf{\textit{I}}) &=&
\sqrt{X^2+Y^2},~~
\vartheta^{l}_{m}(\textbf{\textit{I}}) =
\arctan \left ({Y \over X} \right ), \\
X &=& \sum_{j=0}^{\infty} u^{m}_j
\Psi^{ml}_j(\textbf{\textit{I}}),~~
Y = \sum_{j=0}^{\infty} v^{m}_j
\Psi^{ml}_j(\textbf{\textit{I}}).
\end{eqnarray}
The action $I_2$ remains a constant parameter over a 
time scale of $1/{\cal O}(\epsilon e^{st})$. 

The angle $w_1$ can become a rotating or librating angle 
depending on the initial value of $\mu^{l}_m(\textbf{\textit{I}})$. 
Stars will be near a mean-motion resonance if $w_1$ librates. 
For simplicity, I introduce the slow angle 
$\theta=w_1+\vartheta^{l}_{m}(\textbf{\textit{I}})-\pi$, which 
casts $\overline {\cal K}(m,l)$ into the form 
\begin{equation}
\overline {\cal K}(m,l) = 
{\cal H}_0(\textbf{\textit{I}})-m\Omega_p I_1 -
\epsilon e^{st} A^{l}_{m}(\textbf{\textit{I}})
\cos \theta. \label{eq:averaged-bar-K-theta}
\end{equation}
The distribution function (DF) $f=f_0+f_1$ is thus conserved
along the trajectories determined by
\begin{eqnarray}
\dot \theta \! &=& \frac{\partial \overline {\cal K}(m,l)}{\partial I_1} =
\! \mu^{l}_m(\textbf{\textit{I}}) \! -\! \epsilon
e^{st} \frac {\partial A^{l}_{m}(\textbf{\textit{I}})}{\partial I_1}
\cos \theta,
\label{eq:averaged-equations-vs-theta-1} \\
\dot I_1 \! &=& -\frac{\partial \overline {\cal K}(m,l)}{\partial \theta} =
\! -\epsilon e^{st} A^{l}_{m}(\textbf{\textit{I}})
\sin \theta. \label{eq:averaged-equations-vs-theta-2}
\end{eqnarray}
These are equations of perturbed orbits bound to the $l$th
Fourier component in the linear regime. According to the 
dynamical mechanism presented in \S\ref{sec:capture-into-resonances}, 
the $l=0$ Fourier component is capable of keeping 
$\mu^{l}_m(\textbf{\textit{I}})\approx {\cal O}\left (
\epsilon e^{st} \right )$, which in turn, can lead to the 
resonant capture of stars in the linear regime once $\theta$ 
begins to evolve slowly. 

I have frozen the (small) exponential factor $\epsilon e^{st}$ 
and plotted the instantaneous isocontours (IICs) of 
$\overline {\cal K}(2,0)$ in Figure \ref{pic:resonance-cavity} 
for the initially circular orbits ($I_2=0$) of mode S2. The
isocontours have been displayed in the coordinate plane of 
$u_1=\sqrt{2I_1}\sin \theta$ and $u_2=\sqrt{2I_1}\cos \theta$ 
where the transformation $(\theta,I_1) \rightarrow (u_1,u_2)$ 
is canonical. Note that the angle $\theta$ is measured clockwise 
with respect to the positive $u_2$-direction. 

The IICs of Figure \ref{pic:resonance-cavity} resemble the 
topology of the celestial three-body problem (e.g., Wisdom 1980; 
Winter \& Murray 1997). Two homoclinic loops that intersect at 
the instantaneous hyperbolic point $h_0$ surround a crescent-like, 
resonant region where the angle $\theta$ librates. The instantaneous 
elliptic point $c_0$ corresponds to stable orbits at exact resonance. 
For a specified $I_2$, the coordinates of $c_0$ and $h_0$ are 
$(\theta,I_1)=[\pi,r_{\pi}(t)]$ and $(\theta,I_1)=[0,r_0(t)]$, 
respectively, where $r_{\pi}(t)$ and $r_0(t)$ are the real roots of  
\begin{equation}
\mu^{0}_m(r_z,I_2) - 
\epsilon e^{st} \frac {\partial A^{0}_{m}(r_z,I_2)}{\partial r_z}
\cos z =0,~~z =0,\pi, \label{eq:stationary-points}
\end{equation}
at a given time $t$. 

To this end, I show that the elliptic point $c_0$ always lies on 
the negative $u_2$-axis with $\theta=\pi$. I assume the small 
variations $\tilde \theta=\theta-\pi$ and $\tilde I_1=I_1-r_{\pi}(t)$
for $l=0$, and linearize equations (\ref{eq:averaged-equations-vs-theta-1}) 
and (\ref{eq:averaged-equations-vs-theta-2}) to obtain
\begin{eqnarray}
\frac {d\tilde \theta}{dt} &=& 
a_{12}(t) \tilde I_1,
\label{eq:linearized-equations-vs-theta-1} \\
\frac {d\tilde I_1}{dt} &=& 
a_{21}(t) \tilde \theta -\dot r_{\pi}(t),
\label{eq:linearized-equations-vs-theta-2}
\end{eqnarray}
where the time-dependent coefficients are defined as
\begin{eqnarray}
a_{12}(t) &=& \left [ 
        \frac {\partial \mu^{0}_m(\textbf{\textit{I}})}
              {\partial I_1} 
	      +\epsilon e^{st}
        \frac {\partial^2 A^{0}_m(\textbf{\textit{I}})}
              {\partial I_1^2}
	      \right ]_{I_1=r_{\pi}(t)}, \\
a_{21}(t) &=& \epsilon e^{st} \left [ 
A^{0}_{m}(\textbf{\textit{I}}) \right ]_{I_1=r_{\pi}(t)}.
\end{eqnarray}
The quantity $\dot r_{\pi}(t)$ is determined by 
differentiating (\ref{eq:stationary-points}) with respect 
to $t$. I obtain 
\begin{eqnarray}
\dot r_{\pi}(t) &=& -\frac {a_{13}(t)}{a_{12}(t)}, \\
a_{13}(t) &=& \epsilon s e^{st} 
  \left [  \frac {\partial A^{0}_m(\textbf{\textit{I}})}
           {\partial I_1} \right ]_{I_1=r_{\pi}(t)}.
\end{eqnarray}
Eliminating $\tilde I_1$ from the linearized equations 
leads to  
\begin{eqnarray}
&{}& \frac {d^2\tilde \theta}{dt^2} +
a_{0}(t) \tilde \theta =a_{13}(t)+{\cal O}
\left (\epsilon^2 e^{2st} \right ), 
\label{eq:2nd-order-tilde-theta} \\
&{}& a_{0}(t)= - \left [ 
        \epsilon e^{st} A^{0}_{m}(\textbf{\textit{I}})
        \frac {\partial \mu^{0}_m(\textbf{\textit{I}})}
              {\partial I_1} 
\right ]_{I_1=r_{\pi}(t)}.
\label{eq:spring-coefficient}
\end{eqnarray}
The sign of the time-dependent {\it spring coefficient} $a_0(t)$ 
is determined by the sign of $\partial \mu^0_m/\partial I_1$. In 
cored stellar disks whose equilibrium potential fields are monotonic 
functions of $R$, the following conditions always hold
\begin{equation}
\frac {\partial \Omega_{R}(\textbf{\textit{J}})}{\partial J_{\phi}}\le 0,~~
\frac {\partial \Omega_{\phi}(\textbf{\textit{J}})}{\partial J_{\phi}}\le 0.
\label{eq:negative-diff-omgs-d-I1}
\end{equation}
The equality sign corresponds to the galactic center. The isochrone,
Kuzmin-Toomre and cored exponential disks (investigated in JH and
Paper I) fulfill (\ref{eq:negative-diff-omgs-d-I1}), which implies 
$\partial \mu^0_m/\partial I_1<0$. Therefore, the spring coefficient 
$a_0(t)$ is positive and the homogeneous solution of 
(\ref{eq:2nd-order-tilde-theta}) is bounded. This proves that the 
instantaneous critical point $(\theta,I_1)=[\pi,r_{\pi}(t)]$ is of 
elliptic type. 

The resonant zone disappears for $l\not =0$ when the pattern speed 
lies in the interval (\ref{eq:constraint-pattern-speed-unstable}). 
This is because all participating stars in the pattern formation 
satisfy (\ref{eq:condition-secular-resonance}) that yields
\begin{equation}
\mu^{l}_m(\textbf{\textit{I}}) \approx l\Omega_R+
{\cal O}\left (\epsilon e^{st} \right ).
\label{eq:mu-for-other-l-near-SR}
\end{equation}
Accordingly, $\mu^l_m(\textbf{\textit{I}})$ cannot flip 
sign for $l\not =0$ and equation (\ref{eq:stationary-points}) 
will not have any real root for small perturbations 
of ${\cal O}(\epsilon e^{st})$. In such a circumstance, 
IICs will constitute a bundle of unidirectional closed 
curves that encircle the origin. 
In \S\ref{sec:modes-in-frequency-space}, I will verify for 
the spiral mode S2 that pattern stars satisfy  
(\ref{eq:mu-for-other-l-near-SR}).

\subsection{The Expansion of Resonant Zones}
\label{sec:resonant-zones-expand}

It is evident that the structure of IICs in 
Figure \ref{pic:resonance-cavity} evolves with time as 
the perturbations grow proportional to $\epsilon e^{st}$. 
For small perturbations the critical points $c_0$ and $h_0$ 
are preserved although they are displaced. As a consequence, 
there always exists a resonant zone for $\epsilon e^{st}\ll 1$ 
and it would be interesting to learn how that zone {\it expands} 
in a growing mode. To answer this question analytically, 
I measure the variation of the quantity 
\begin{equation}
\Delta = d_{o}-d_{i},
\end{equation}
where $d_{o}$ and $d_{i}$ are, respectively,
the values of $I_1$ on the outer and inner homoclinic loops at 
$\theta=\pi$. The implicit functional form of homoclinic 
loops is given by
\begin{eqnarray}
&{}& \overline {\cal K}_0(I_1,I_2) -
\epsilon e^{st} A^{0}_{m}(I_1,I_2) \cos \theta =\nonumber \\
&{}& \qquad \overline {\cal K}_0[r_0(t),I_2] -
\epsilon e^{st} A^{0}_{m}[r_0(t),I_2],
\end{eqnarray}
whose temporal derivative at $\theta=\pi$ results in
\begin{eqnarray}
&{}& \!\! \dot d_{\nu} \left [ \mu^0_m +
\epsilon e^{st} 
\frac {\partial A^{0}_{m}}{\partial I_1}
\right ]_{I_1=d_{\nu}} \!\!\!\!\! + 
\epsilon s e^{st} A^{0}_{m}(d_{\nu},I_2) 
= \nonumber \\
&{}& \!\! \dot r_0 \left [ \mu^0_m \!-\!
\epsilon e^{st} 
\frac {\partial A^{0}_{m}}{\partial I_1}
\right ]_{I_1=r_0} \!\!\!\!\! - 
\epsilon s e^{st} A^{0}_{m}(r_0,I_2),
 \label{eq:growth-of-Io-and-Ii}
\end{eqnarray}
for $\nu \equiv o,i$. The bracket on the right hand side of 
equation (\ref{eq:growth-of-Io-and-Ii}) vanishes because of 
(\ref{eq:stationary-points}) and the bracket on the left 
hand side is simply $d\theta_{\nu}/dt$ ($\nu$=$o,i$). 
From (\ref{eq:averaged-equations-vs-theta-1}) and 
(\ref{eq:averaged-equations-vs-theta-2}) one can 
verify that $d\theta_o/dt<0$ and $d\theta_i/dt>0$.
Equation (\ref{eq:growth-of-Io-and-Ii}) thus becomes 
\begin{equation}
\dot d_{\nu} = - \frac {\epsilon s e^{st} }
                       {\dot \theta_{\nu} } 
\left [ A^{0}_{m}\left (d_{\nu},I_2 \right ) 
+ A^{0}_{m} \left (r_0,I_2 \right ) \right ],~~
\nu \equiv o,i. \label{eq:growth-of-Io-and-Ii-closed-form}
\end{equation}
Subtracting $\dot d_i$ from $\dot d_o$ leads to
\begin{eqnarray}
\dot \Delta &=& s \left ( \epsilon e^{st} \right )
   \Biggl [ 
       \frac {A^{0}_{m}\left (d_o,I_2 \right )}{|\dot \theta_o|} +
       \frac {A^{0}_{m}\left (d_i,I_2 \right )}{|\dot \theta_i|} 
       \nonumber \\ 
     &{}& \qquad \qquad + \frac {A^{0}_{m} \left (r_0,I_2 \right )}
	        {|\dot \theta_o|} +
          \frac {A^{0}_{m} \left (r_0,I_2 \right )}
	        {|\dot \theta_i|} \Biggr ].
\label{eq:growth-of-Delta}
\end{eqnarray}
The angular rates $\dot \theta_o$ and $\dot \theta_i$ 
can be estimated using the linearized equation 
(\ref{eq:linearized-equations-vs-theta-1}) as
\begin{equation}
\dot \theta_o=a_{12}(t) \left [ d_o-r_{\pi}(t) \right ],~~
\dot \theta_i=a_{12}(t) \left [ d_i-r_{\pi}(t) \right ].
\label{eq:relation-for-dot-theta_o-and-theta_i}
\end{equation}
Since the resonant zone is thin, one may assume 
$r_{\pi}(t)\approx \left ( d_o +d_i \right )/2$, 
which can be combined with 
(\ref{eq:relation-for-dot-theta_o-and-theta_i})
to obtain
\begin{equation}
|\dot \theta_o | \approx 
|\dot \theta_i | \approx 
|a_{12}(t) | \frac{\Delta}{2}.
\label{eq:approx-relation-for-dot-theta_o}
\end{equation}
Substituting from (\ref{eq:approx-relation-for-dot-theta_o})
in (\ref{eq:growth-of-Delta}) yields
\begin{eqnarray}
\Delta \dot \Delta & \approx & 
\frac{2 s \left ( \epsilon e^{st} \right )}{|a_{12}(t) |} 
   \Bigl [ 
       2 A^{0}_{m}\left (r_0,I_2 \right ) \nonumber \\ 
     &{}&  \qquad + 
       A^{0}_{m}\left (d_i,I_2 \right ) + 
       A^{0}_{m}\left (d_o,I_2 \right ) \Bigr ],
\label{eq:growth-of-Delta-2nd-relation}
\end{eqnarray}
where the quotient of the barcketed terms and $|a_{12}(t)|$ is 
of ${\cal O}(1)$. Integrating (\ref{eq:growth-of-Delta-2nd-relation}) 
results in
\begin{equation}
\Delta \approx 2 \sqrt{ \epsilon e^{st} } {\cal O}(1),~~
\dot \Delta \approx s \sqrt{\epsilon e^{st} } {\cal O}(1),
\label{eq:growth-of-Delta-and-dotDelta}
\end{equation}
from which the expansion rate of the resonance width per unit 
length is estimated:
\begin{equation}
\dot \Delta/\Delta \sim s/2.
\label{eq:growth-of-Delta-versus-s}
\end{equation}
This shows that stars are steadily captured into resonance 
by the $l=0$ Fourier component as the mode grows in the linear 
regime. 

My calculations show that the maximum of $A^0_m(\textbf{\textit{I}})$ 
in unstable modes does not coincide neither with $h_0$ nor with $c_0$, 
but it falls inside the resonant zone within a time scale  
$\ll 1/{\cal O}\left (\epsilon e^{st} \right )$ as $\Delta$
grows from zero width. The observed shift from the location 
of $c_0$ depends on the gradient $\partial f_0/\partial J_{\phi}$. 
All growing modes obey the following rules near $c_0$:
\begin{eqnarray}
&{}& 
\frac{\partial A^0_m(\textbf{\textit{I}})}{\partial I_1}
<0<\mu^0_m(\textbf{\textit{I}})~~{\rm if}~~ 
\frac{\partial f_0(\textbf{\textit{J}})}{\partial J_{\phi}}<0, 
\label{eq:negative-gradient-of-A} \\
&{}& 
\frac{\partial A^0_m(\textbf{\textit{I}})}{\partial I_1}
>0>\mu^0_m(\textbf{\textit{I}})~~{\rm if}~~ 
\frac{\partial f_0(\textbf{\textit{J}})}{\partial J_{\phi}}>0.
\label{eq:positive-gradient-of-A}
\end{eqnarray}
In both cases the magnitude of $\mu^0_m(\textbf{\textit{I}})$ 
remains small near the maximum of $A^0_m(\textbf{\textit{I}})$, 
confirming the theory of \S\ref{sec:capture-into-resonances}
that the perturbed density profile is supported by the 
synchronous precession of orbital axes.

\begin{figure*}
\centerline{\hbox{\includegraphics[width=0.9\textwidth]
{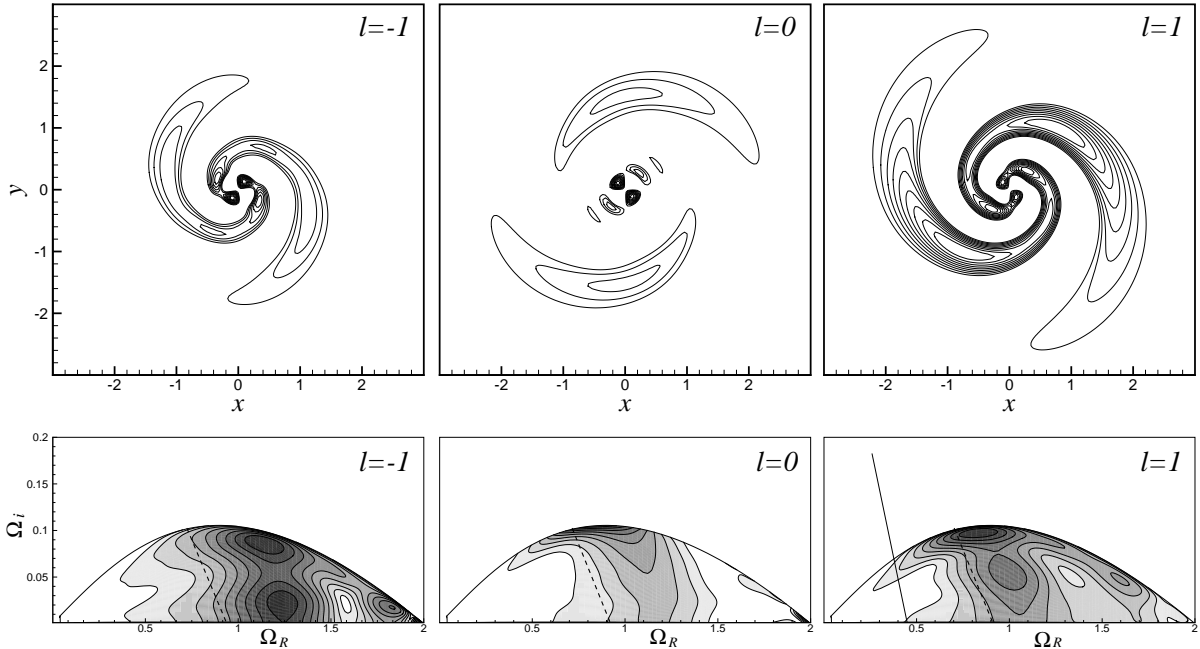} }} \caption{The density components
$\Sigma^{l}_1$ (top panels) and the contours of $A^l_2({\bf
\Omega})$ (bottom panels) for mode S2. In all figures dashed lines
mark the stars at exact CR with $\mu^0_2({\bf \Omega})=0$. The line
$\mu^{-1}_2({\bf \Omega})=0$ lies outside the plot range because it
is parallel to the $\Omega_R$-axis with $\Omega_i=0.454$. Straight
solid line in the bottom-right panel is defined by $\mu^1_2({\bf
\Omega})=0$. Its relative location with respect to the highly
populated regions of the frequency space shows that stars bound to
the $l=1$ component evolve far from the OLR. The contour levels of
$A^l_2({\bf \Omega})$ range from 10\% to 90\% of the maximum with
increments of 10\%. \label{pic:components-and-zones-B1-S2}}
\end{figure*}

\subsection{Angular Momentum Transfer}
\label{sec:angular-momentum-at-resonance}

The location of the resonant zone in the $(u_1,u_2)$-plane 
provides valuable information of the behavior of resonant 
stars. The Hamiltonian varies according to the equation 
\begin{equation}
\frac {d \overline {\cal K}(m,l)}{dt}=
\frac {\partial \overline {\cal K}(m,l)}{\partial t}=
-\epsilon s e^{st} A^l_m(\textbf{\textit{I}}) \cos \theta.
\label{eq:formula-for-bar-K-variation}
\end{equation}
One can thus determine how the orbital energy of trapped stars 
changes over time. Recalling that the instantaneous elliptic point 
$c_0$ occurs at $\theta=\pi$, for most resonant orbits the 
minimum of $\theta$ is larger than $\pi/2$ 
(see Figure \ref{pic:resonance-cavity}) and equation 
(\ref{eq:formula-for-bar-K-variation}) yields
\begin{equation}
\frac {d \overline {\cal K}(m,0)}{dt}>0.
\end{equation}
In other words, stars gain energy during a librational 
motion around $\theta=\pi$. Since $I_2=J_R$ (for $l=0$) 
is an adiabatic invariant in the resonance zone, an 
increase in the orbital energy of stars boosts their 
angular momentum. 

There is a different mechanism for the angular momentum transfer 
to/from stars whose $\theta$ is rotational. Such stars will definitely
experience both $\dot I_1>0$ and $\dot I_1<0$ states over a complete
period of $\theta$, but which one overwhelms the other and how does it 
affect the orbital angular momentum? Outside the resonant zone we have 
\begin{equation}
\left \vert \mu^l_m(\textbf{\textit{I}}) \right \vert 
\gg \left \vert \epsilon e^{st} 
\frac {\partial A^{l}_{m}(\textbf{\textit{I}})}{\partial I_1} 
\right \vert,
\end{equation}
which guarantees the rotation of $\theta$ according 
to (\ref{eq:averaged-equations-vs-theta-1}). I now divide 
equation (\ref{eq:averaged-equations-vs-theta-2}) by 
(\ref{eq:averaged-equations-vs-theta-1}) and ignore 
all terms of ${\cal O}(\epsilon^2 e^{2st})$ to obtain 
\begin{equation}
\frac {dI_1}{d\theta}=-\frac {\epsilon}{\mu^l_m(\textbf{\textit{I}}_0)}
\exp \left [\frac{s\theta}{\mu^l_m(\textbf{\textit{I}}_0)} \right ]
A^l_m(\textbf{\textit{I}}_0)\sin \theta,
\label{eq:I1-vs-theta}
\end{equation}
where $\textbf{\textit{I}}_0$ is the initial value of the 
action vector $\textbf{\textit{I}}=(I_1,I_2)$ at $\theta=0$. 
For a clockwise rotation of phase space flows, $\theta$ ranges 
from $0$ to $2\pi$ and reversely for their counter-clockwise 
rotation. Moreover, the direction of rotation is determined by 
the sign of $\mu^l_m(\textbf{\textit{I}}_0)$. 
Given these points, integrating (\ref{eq:I1-vs-theta}) over 
a complete cycle of $\theta$ results in the incremental 
change of $I_1$ as
\begin{equation}
\Delta I_1(l)=\epsilon A^l_m 
\frac {\mu^l_m
       {\rm sign}\left (\mu^l_m \right )}
       {\left ( \mu^l_m \right )^2+s^2}
  \left [ \exp \left ( \frac {2\pi s}{\mu^l_m} \right ) -1
  \right ].    
  \label{eq:delta-I1-vs-l} 
\end{equation}
It is seen that $\Delta I_1$ is positive for 
$\mu^l_m(\textbf{\textit{I}}_0)>0$ and stars gain 
angular momentum as the perturbations grow. The opposite 
happens for $\mu^l_m(\textbf{\textit{I}}_0)<0$. Combining 
(\ref{eq:mu-for-other-l-near-SR}) and (\ref{eq:delta-I1-vs-l})
shows that the angular momentum gain or loss is decided 
by the sign and magnitude of the Fourier number $l$. The Fourier 
components with $l<0$ and $l>0$ respectively drain and boost the 
angular momentum of stars so that 
\begin{eqnarray}
&{}& \!\! \Delta I_1(-1)<\Delta I_1(-2)<
\Delta I_1(-3)<\cdots <0, \label{eq:sort-negative-l} \\
&{}& \!\! \Delta I_1(+1)>\Delta I_1(+2)>
\Delta I_1(+3)>\cdots >0. \label{eq:sort-positive-l}
\end{eqnarray}

Not all stars bound to the $l=0$ component are in the 
resonant zone, specially in the limit of 
$\epsilon e^{st}\rightarrow 0$. Hence, the sign of 
$\Delta I_1(0)$ is determined both by resonant and 
non-resonant stars. Based on my arguments presented 
after equation (\ref{eq:formula-for-bar-K-variation}), 
resonant stars always gain angular momentum and have a 
positive contribution to $\Delta I_1(0)$. However, It is
the sign of $\partial f_0/\partial J_{\phi}$
that determines whether non-resonant stars of the $l=0$
component are emitting or absorbing angular momentum. 
Equations (\ref{eq:negative-gradient-of-A}) and 
(\ref{eq:positive-gradient-of-A}) together with 
(\ref{eq:delta-I1-vs-l}) show that non-resonant stars
of the $l=0$ component gain and lose angular momentum for 
$\partial f_0/\partial J_{\phi}<0$ and 
$\partial f_0/\partial J_{\phi}>0$, respectively. 
Therefore, $\Delta I_1(0)$ is certainly positive if the 
mode develops in a region of the phase space with 
$\partial f_0/\partial J_{\phi}<0$. Since most stars of 
the $l=0$ component are non-resonant as 
$\epsilon e^{st}\rightarrow 0$, the occurrence of a 
growing mode as a result of $\partial f_0/\partial J_{\phi}>0$ 
will lead to $\Delta I_1(0)<0$. The results of the above 
analysis are consistent with the results of JH and the 
bar charts of Figure \ref{pic:ang-momentum-modes-B1-S2}.

\subsection{Mode Components in the Frequency Space}
\label{sec:modes-in-frequency-space}

To understand how the mechanism of resonant trapping operates
on non-circular orbits in the frequency space, I have plotted 
the components of mode S2 in Figure \ref{pic:components-and-zones-B1-S2}. 
Top panels display the pattern components in the configuration space, 
and in bottom panels I have shown the isocontours of $A^l_2({\bf \Omega})\equiv
A^l_2(\textbf{\textit{I}})$ together with the lines $\mu^1_2({\bf
\Omega})=0$ (straight solid line) and $\mu^0_2({\bf \Omega})=0$
(dashed lines). Darker regions in the bottom panels correspond to
larger values of $A^l_2({\bf \Omega})$. The contour plots of
$A^l_2({\bf \Omega})$ show which stars in the frequency/action space
are engaged with the $l$th component. As one could anticipate, the
highly populated regions of all components are close to the CR zone
of Figure \ref{pic:CR-zones}. The maxima of $A^l_2({\bf \Omega})$
have not been located on the line $\mu^0_2({\bf \Omega})=0$
because the initial density gradient of the equilibrium state
displaces the center of mass of pattern stars. In modes with
$\partial f_0/\partial J_{\phi}<0$ near the CR zone, the maxima 
are shifted to regions with $\mu^0_2({\bf \Omega})>0$ 
(see \S\ref{sec:resonant-zones-expand}). 

Whilst the $l=0$ component has trapped near-circular orbits into the
CR (Figure \ref{pic:components-and-zones-B1-S2}), the stars bound to
other components evolve far from mean-motion resonances and the
angle $\theta$ becomes rotational for them. In fact, stars bound 
to the $l=-1$ component avoid the line $\mu^{-1}_m({\bf \Omega})=0$ 
and the quantity $\mu^{-1}_m({\bf \Omega})$ is always negative as long 
as inequality (\ref{eq:condition-for-NO-ILR}) holds. Through a similar 
mechanism, stars avoid the line $\mu^{1}_m({\bf \Omega})=0$ and the 
quantity $\mu^{1}_m({\bf \Omega})$ remains positive regardless of the 
magnitude of $\Omega_p$. These results show that the dynamical mechanism 
for growing modes of finite $s$ differers with LBK's theory established 
in the limit of $s\rightarrow 0$. In other words, in a growing spiral
mode similar to mode S2 of this paper, the $l=-1$ and $l=1$ components 
are not associated with the ILR and OLR. A new dynamical origin of 
instabilities, which interprets the results of this paper and has not 
the restrictions of LBK's theory, is presented in the next section.

\section{THE ORIGIN OF INSTABILITIES}
\label{sec:origin-of-instabilities}

Consider a randomly generated, small-amplitude, $m$-fold density wave 
of pattern speed $\Omega_p$ whose potential field can be generally 
expanded in the Fourier series of the angle variables. Independent 
of the form of initial phase space distribution, determined by 
$f_0(\textbf{\textit{J}})$, the points 
$\left (\textbf{\textit{I}},\theta \right )=
\left (\textbf{\textit{I}}_0, 0\right )$ and 
$\left (\textbf{\textit{I}},\theta \right )=
\left (\textbf{\textit{I}}_0, \pi \right )$ will emerge as 
time-invariant stationary points of the flows generated by 
$\overline {\cal K}(m,0)$ if the following condition holds
\begin{equation}
\frac {\partial A^0_m(\textbf{\textit{I}})}{\partial I_1}
=\mu^0_m(\textbf{\textit{I}})=0,
\label{eq:condition-for-stationary-points}
\end{equation}
at $\textbf{\textit{I}}=\textbf{\textit{I}}_0$.
In such a circumstance, orbits associated with either of these 
stationary points will never change their action vector 
$\textbf{\textit{I}}=(I_1,I_2)$ and their orbital energy must be 
conserved. This condition will be satisfied only if $s=0$ in 
(\ref{eq:formula-for-bar-K-variation}). This is how a \citet{vK55} 
mode is born. The state of (\ref{eq:condition-for-stationary-points}) 
can occur at any point in the infinite dimensional action space 
and for arbitrary values of $\Omega_p$. Consequently, stationary 
modes constitute a continuous family and Mathur's (1990) isolated, 
pure oscillatory modes are not feasible in stellar disks.

Let me now suppose that for some pattern speed $\Omega_p$ in 
the interval (\ref{eq:constraint-pattern-speed-unstable}), the 
quantity $\mu^0_m(\textbf{\textit{I}})$ accidentally remains 
small (for a group of stars) over a finite duration of time,
but the condition (\ref{eq:condition-for-stationary-points}) 
is violated. This (likely) symmetry-breaking phenomenon creates 
a slim resonant zone of the width $\Delta(t)$ (see 
\S\ref{sec:resonant-zones-expand}). There is not any 
physical constraint on the evolution of $\Delta(t)$ when 
the perturbations are in their early stages. For a steady 
rate given by (\ref{eq:growth-of-Delta-versus-s}) more stars 
are trapped by the resonant zone if $s>0$ and the density 
wave corresponding to the Fourier number $l=0$ is magnified. 
As I discussed in \S\ref{sec:angular-momentum-at-resonance}, 
the angular momentum of the captured stars increases and some 
other {\it reacting stars} should therefore respond in order 
to recover the angular momentum balance of the disk. 
The reacting stars live in the same CR zone defined by 
$\mu^0_m({\bf \Omega})\approx {\cal O}
\left (\epsilon e^{st} \right )$. 

As more stars are trapped into resonance by the $l=0$ 
component, non-resonant components with $\vert l\vert >0$ 
should also involve more stars to compensate the angular 
momentum deficit. So the amplitudes of other wave components 
increase as well. This is what I am considering as the 
origin of instabilities: triggering unsteady density 
perturbations by the capture of stars into the CR. 
Angular momentum transfer between non-resonant 
stars cannot be facilitated if a resonant gap does not open 
in the phase space. In fact, the initial destabilizing 
imbalance of ${\cal L}$ is generated by an irreversible 
engagement of resonant stars even if the amount of angular 
momentum that they absorb is small (e.g., Figure 8 in JH). 
A perturbation with the property of 
(\ref{eq:condition-for-stationary-points}) fixes 
$s=0$, does not awake stars of $\vert l\vert >0$ components,
and it creates a stationary mode. I conclude that most 
(if not all) unstable modes disappear for $\Omega_p>\Omega_0$ 
just because the CR is destroyed in such a circumstance.

The $l=0$ component is not always an angular momentum absorber. 
For example, in models with an inner cutout of the DF, one has 
$\partial f_0/\partial J_{\phi}>0$ which enforces $L_m(0)<0$ 
(discussions in \S\ref{sec:angular-momentum-at-resonance} and JH). 
Figures 2 and 11 in JH demonstrate two modes that show angular
momentum emission by the $l=0$ component. The pattern speeds 
of some C-modes explored in Paper I exceed $\Omega_0$ marginally.  
Those modes also have $L_m(0)<0$. Unstable modes with 
$\Omega_p>\Omega_0$ are barely observed in modal calculations 
because their large pattern speeds shrink them to the galactic 
center. Coexistence of mode B1 and a compact C-mode can explain 
the origin of double-barred galaxies. For a softened-gravity 
model of the exponential disk with an inner cutout, Alar Toomre 
(private communication) also finds an inner edge mode that has 
a corotation circle inside its pattern. Toomre's inner edge mode 
is another example of a mode with $\partial f_0/\partial J_{\phi}>0$ 
and $L_m(0)<0$. Axisymmetric features/grooves introduced by 
\citet{SK91} to the surface density and DF also develop a spiral 
mode whose pattern is discontinuous at the location of the groove.
That discontinuity is due to the rapid change in the sign of 
$\partial f_0/\partial J_{\phi}$: while the $l=0$ component 
boosts the angular momentum of stars outside the groove, 
it has an opposite effect on inner stars.

Each unstable mode occupies a finite region in the frequency space
due to the finiteness of its CR zone. This is how divisions (gaps)
are created between the pattern speeds of unstable modes, the ${\bf
\Omega}$-space is quantized by resonances, and finally, unstable
modes constitute a discrete family. Growth rates are sorted according 
to the likelihood of resonant capture, which is proportional to the 
initial population of stars determined by the equilibrium DF. In a 
stellar disk with a falling density profile, the growth rates of 
linear modes are sorted in a decreasing order from compact modes 
to more extensive ones. Most barred and spiral modes obey this 
rule (see Paper I). The perturbation theory of 
\S\ref{sec:perturbed-stellar-dynamics} says that stars live 
on the integral manifolds of constant $I_2$ before and after 
capture into resonance. In order to guarantee a sustained growth 
of the CR zone, the initial DF $f_0(\textbf{\textit{J}})$ must 
be smooth and non-zero along the curves of constant $I_2$, 
and $\mu^{0}_m(\textbf{\textit{I}})$ has to flip its sign 
there. A resonant zone cannot grow in a region of the phase space 
which is devoid of stars from the beginning, or its stars have been 
forced to migrate by unsteady processes. The first-order perturbation 
theory is valid as long as the CR zone of each mode is inaccessible 
by the stars of other modes and $I_2$ is an adiabatic invariant. 

\begin{figure}
\centerline{\hbox{\includegraphics[width=0.45\textwidth]
{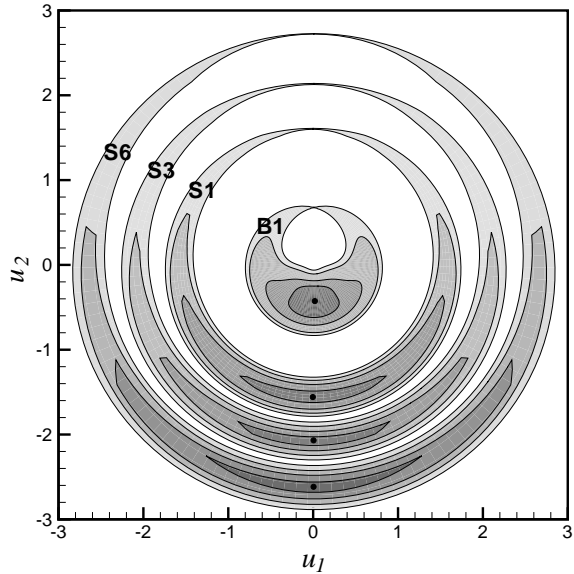} } }
\caption{The resonant zones of modes B1, S1, S3 and S6 for 
the cored exponential disk with $(N,\lambda,\alpha)=(6,1,0.42)$.
Contour plots show the IICs of $\overline {\cal K}(2,0)$ for 
$I_2=0$ and $\epsilon e^{st}=0.01$. Darker regions
correspond to larger values of $\overline {\cal K}(2,0)$. 
Contour levels of each zone have been normalized to 
the maximum value of $\overline {\cal K}(2,0)$ in the 
same zone. 
\label{pic:zones-B1-S1-S3-S6}}
\end{figure}
\begin{figure*}
\centerline{\hbox{\includegraphics[width=0.45\textwidth]
             {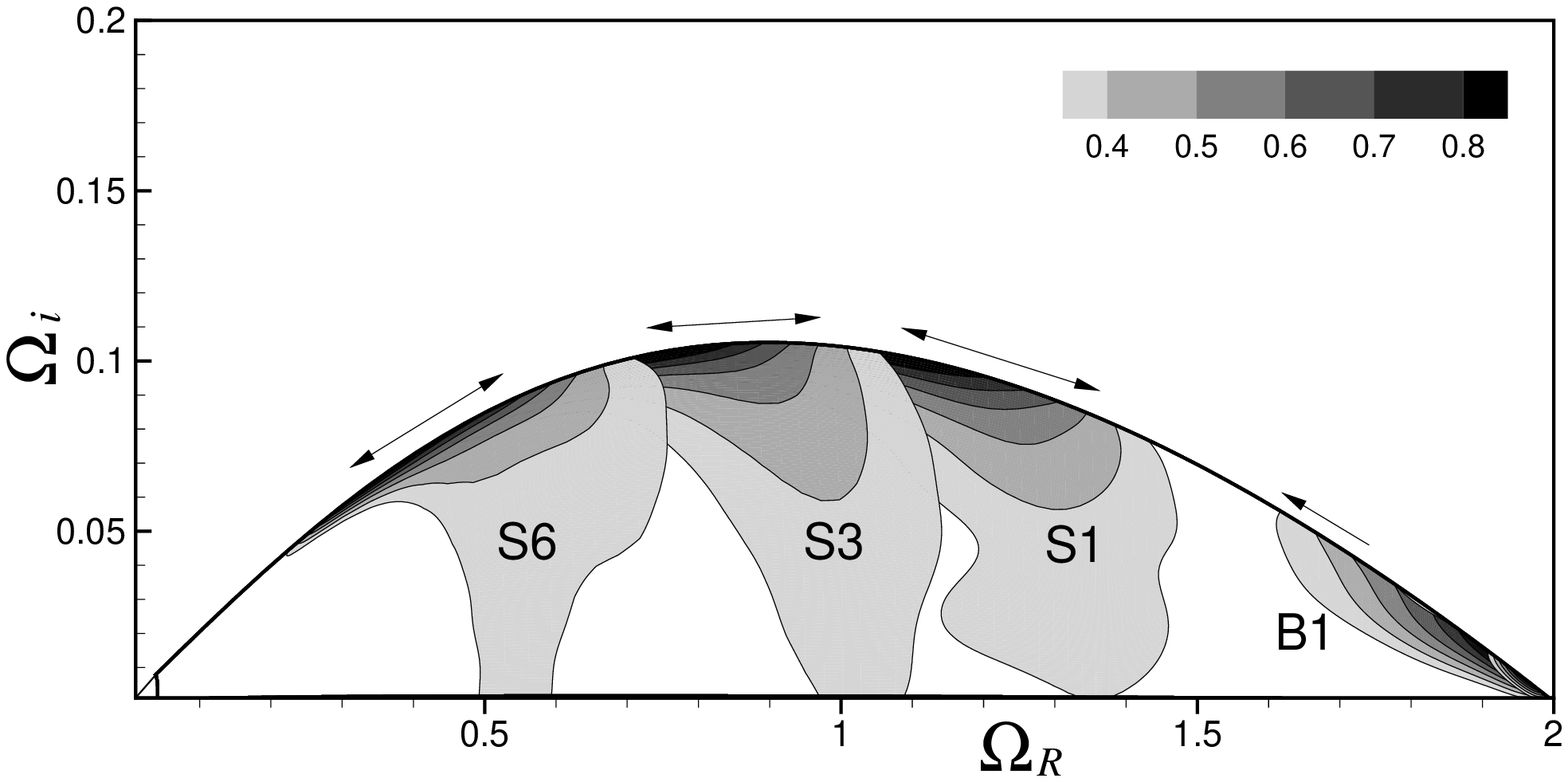}}
            \hbox{\includegraphics[width=0.45\textwidth]
             {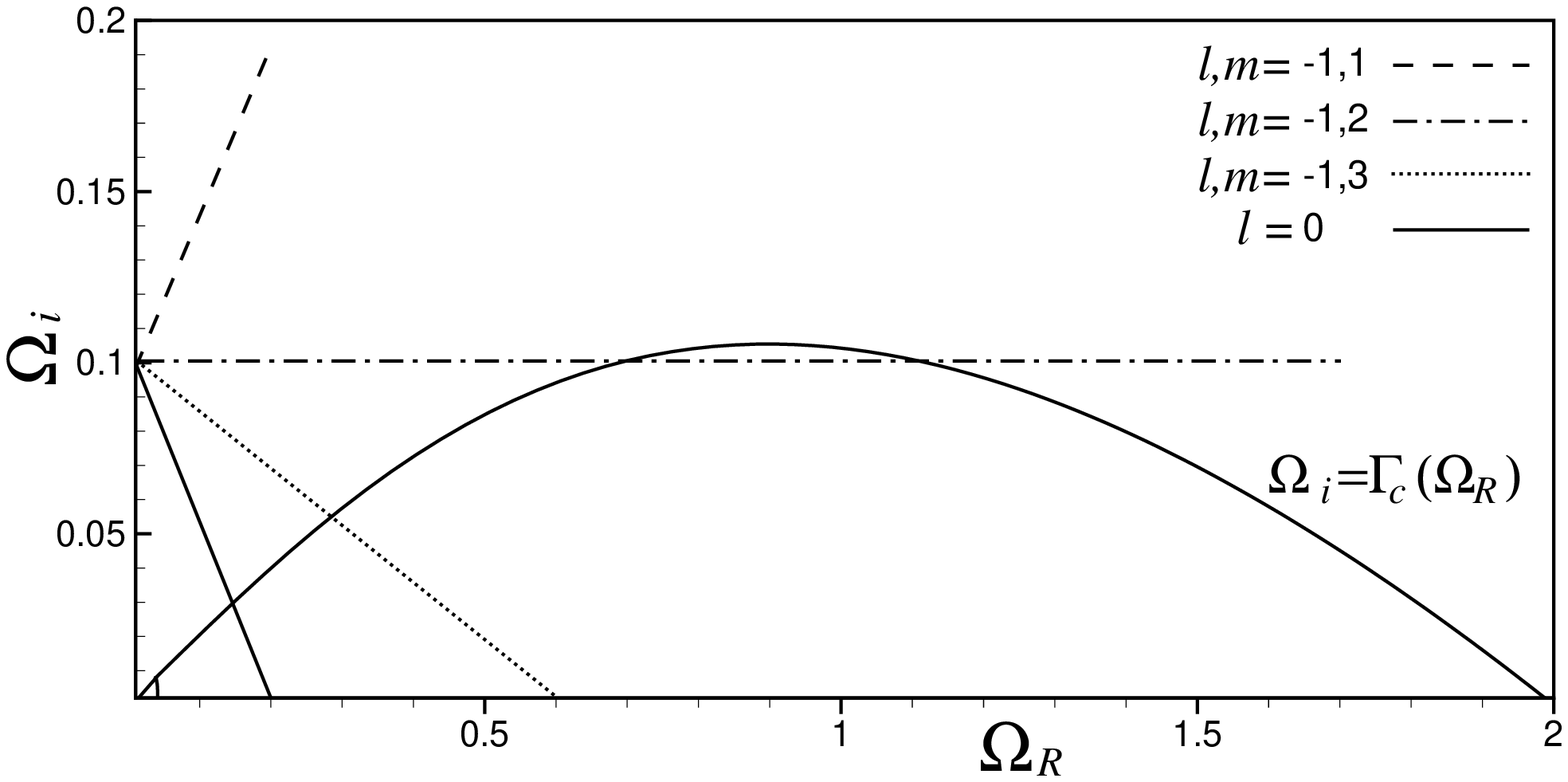}} }
\caption{{\it Left panel}: Contour plots of $A^0_2({\bf \Omega})$
for modes B1, S1, S3 and S6 of the cored exponential disk with 
$(N,\lambda,\alpha)=(6,1,0.42)$. The levels of isocontours have 
been normalized to the maximum amplitude. {\it Right panel}: 
The emergence of ILR through the intersection of the lines 
$\mu^{-1}_m({\bf \Omega})=0$ with the frequency space. 
The line $\mu^0_m({\bf \Omega})=0$ (it is independent
of $m$) has also been plotted to show the location of the CR.
In all cases the pattern speed has been set to $\Omega_p=0.1$.
There is no ILR associated with $m=1$ perturbations for
any $\Omega_p>0$. \label{pic:zones-overlap}}
\end{figure*}

\section{MODE SATURATION AND STABLE DISKS}
\label{sec:mode-saturation}

Perhaps the most important issue related to the growth of
instabilities is that the amplitudes of unstable modes are 
saturated after several pattern rotations. Saturation 
is indeed a nonlinear phenomenon and can have complex routes. 
\citet{SB02} suggest that the growth of a mode is stopped by 
the emergence of horseshoe orbits. What they call 
``horseshoe orbits" are simply the librational orbits captured 
by the CR. I found those orbits the main cause of instabilities. 
I agree with \citet{SB02} that the linear perturbation theory 
must fail at some stage, but it does not necessarily imply 
that resonant zones can self-control their expansion beyond 
the linear regime. Using the phase space geometry and the 
nonexistence of the first integral $I_2$ at the boundaries 
of neighboring resonances, I argue that resonance overlapping 
is an alternative and efficient mechanism to stop the growth of 
unstable modes. The stabilization of stellar disks in the 
presence of the ILRs is also explained by the same mechanism. 
I then present a simple criterion for the stability of 
non-axisymmetric perturbations, and discuss about the 
stability of self-consistent scale-free disks.

\subsection{Resonance Overlapping} 
\label{sec:resonance-overlap}

As a mode grows, the width of its CR zone increases proportional 
to $\sqrt{\epsilon e^{st}}$ until two adjacent zones, corresponding 
to modes of different pattern speeds, overlap and compete for trapping 
the stars in the overlapping zone. This is the moment that a chaotic 
layer occurs in the phase space according to Chirikov's (1979) theory, 
and stars living in that layer repeatedly migrate between different 
resonant zones. Thus, the share contributed to each competing resonant 
zone from stars in the overlapping region drops substantially and the
growth is suppressed. A natural consequence of the emergence of a
chaotic layer is the radial migration of stars that heats up the
disk. Some modes may be gradually dissolved as their
Kolmogorov-Arnold-Moser (KAM) tori around stable periodic orbits are
destroyed by modes with larger resonant zones and growth rates. I
remark that the weighted residual form of the CBE that I obtained in
Paper I is valid as long as we are allowed to average out angle
variables from dynamical equations. After the occurrence of chaotic
orbits it is impossible to average out resonant angles and extra
coupling terms between the amplitudes of unstable modes are
contributed to the reduced CBE. Details of my numerical 
simulations in the nonlinear regime and in the presence of 
chaotic orbits will be presented in Paper III.

Figure \ref{pic:zones-B1-S1-S3-S6} shows the $l=0$ resonant 
zones of modes B1, S1, S3, and S6 for the initially circular 
orbits ($I_2=0$) of the model introduced in \S\ref{sec:used-model}. 
This is the first demostartion of its kind that shows how the 
resonant zones of different unstable modes partition the phase 
space of a stellar disk. Since the perturbations grow according 
to an exponential law, resonant zones had been slim rings infinitely 
long ago. The structure shown in Figure \ref{pic:zones-B1-S1-S3-S6} 
has been obtained by setting $\epsilon e^{st}=0.01$ for all modes. 
Correspondingly, resonant gaps are of comparable size at the 
displayed moment but they will not remain so because the growth 
rates are different. Left panel in Figure \ref{pic:zones-overlap} 
shows the amplitude function $A^0_2({\bf \Omega})$ of the same 
modes of Figure \ref{pic:zones-B1-S1-S3-S6} but this time in the 
frequency space that covers all possible values of $I_2$. 
Arrows indicate the expansion direction of the $l=0$ Fourier
component as the modes grow. The growth rates of modes are 
different and mode S6, which is at the bifurcation point 
of the S-family (see Figure 2{\em a} in Paper I), has the 
smallest $s$. The configuration of the CR zones in the 
phase/frequency space and their thickening according to 
equation (\ref{eq:growth-of-Delta-versus-s}) suggest that 
resonance overlapping is likely in stellar disks with rich 
branches of unstable modes in their eigenspectra. 

To find out how and when an overlap takes place, I assume 
two growing waves with the eigenvalues $m\Omega_{p,1}+\imath s_1$ 
and $m\Omega_{p,2}+\imath s_2$ that successively appear in the 
spectrum with $\Omega_{p,1}>\Omega_{p,2}$. These waves develop 
CR zones of widths $\Delta_1(I_2)$ and $\Delta_2(I_2)$ on each 
manifold of constant $I_2$, and I define $d_{12}(I_2)$ as the 
distance between the instantaneous elliptic points of those zones. 
Let me suppose that the global minimum of $d_{12}(I_2)$ corresponds 
to the critical action $I_2=I^{\rm cr}_2$. An overlap thus occurs if 
\begin{equation}
d_{12}(I^{\rm cr}_2) \lessapprox 
\frac 12 \left [ \Delta_1(I^{\rm cr}_2)+
                 \Delta_2(I^{\rm cr}_2) 
         \right ].
\label{eq:condition-for-1st-overlap}
\end{equation}
On the other hand, the magnitude of $d_{12}(I^{\rm cr}_2)$ 
can be estimated as
\begin{equation}
d_{12}(I^{\rm cr}_2)\approx  
I_{1,2}-I_{1,1}+{\cal O}\left (\epsilon e^{st} \right ),
\label{eq:estimate-of-d12}
\end{equation}
where $\Omega_{\phi}\left (I_{1,j},I^{\rm cr}_2 \right)=\Omega_{p,j}$
($j$=1,2). By combining (\ref{eq:condition-for-1st-overlap}) and 
(\ref{eq:estimate-of-d12}), and using 
(\ref{eq:growth-of-Delta-and-dotDelta}), I obtain 
\begin{equation}
I_{1,2} - I_{1,1}  
\lessapprox \left [ \sqrt{\epsilon e^{s_1t}}{\cal O}_1(1)
\!+\!
\sqrt{\epsilon e^{s_2t}}{\cal O}_2(1) \right ]_{I_2=I^{\rm cr}_2},
\label{eq:final-condition-for-overlap}
\end{equation}
which will be satisfied in the linear regime at some $t=t_{\rm cr}$ 
for a sufficiently small value of $I_{1,2}-I_{1,1}$. If the post-linear 
changes in the DF allow for a sustained expansion of resonant zones (when 
the amplitude growth is no longer exponential), one can still anticipate 
an overlap for larger values of $I_{1,2}-I_{1,1}$. Cooling a stellar disk 
increases the number of unstable modes (Figure 6 in Paper I) and decreases 
the gaps between the CR zones. Consequently, transition to chaos is faster 
in cold disks than hot ones. This result is in harmony with Sellwood's 
(2007, private communication) $N$-body experiments. The very first overlapping 
does not necessarily occur between the modes of the same azimuthal wavenumber 
$m$. The pattern speeds of two rapidly growing modes of different $m$ may 
result in a very small $I_{1,2}-I_{1,1}$ and thus cause an early-stage 
overlap. 

The fundamental bar mode may still be saturated in the 
absence of spiral modes and in a shorter time scale than 
chaotic diffusion of stars. When the resonant zone of mode 
B1 grows, the inner homoclinic loop rapidly shrinks to 
central regions, the supply of stars to the CR zone is cut 
from the galactic center, and the bar mode saturates there. 
On the other hand, the bar continues to elongate because 
of the expansion of the outer homoclinic loop. The elongation 
rate, however, will slow down if $\partial f_0/\partial J_{\phi}<0$.
A declining rate of amplitude growth gives time to the trapped 
stars to fill their invariant tori and maintain the bar in 
outer regions.

\subsection{The Role of Inner Lindblad Resonances}
\label{sec:role-of-ILR}

Disappearance of unstable modes when the pattern speed violates
(\ref{eq:condition-for-NO-ILR}) can also be explained by the same
mechanism that saturates instabilities. The ILR emerges when the
line $\mu^{-1}_m({\bf \Omega})=0$ crosses the circular orbit
boundary $\Omega_i=\Gamma_c(\Omega_R)$. This happens while the line
$\mu^0_m({\bf \Omega})=0$ has a crossing too and the CR zone has
already been created. The competition between the ILR and CR near 
their neighboring boundaries can therefore prohibit both resonances
from broadening their zones and the disk develops a stationary wave. 
Right panel in Figure \ref{pic:zones-overlap} shows 
the frequency space of the cored exponential disk together with the 
lines $\mu^{0}_m({\bf \Omega})=0$ and $\mu^{-1}_m({\bf \Omega})=0$ 
that have been drawn for $\Omega_p=0.1$ and $m=$1,2,3. 
The line $\mu^{-1}_1({\bf \Omega})=0$ never intersects the 
frequency space independent of the magnitude of $\Omega_p>0$. 
A crossing is possible for $m\ge 2$ and the slope of 
$\mu^{-1}_m({\bf \Omega})=0$ increases proportional to $m$. 
Subsequently, the distance between the CR and ILR zones decreases 
by increasing $m$ and the disk shows more reluctance against 
developing an unstable mode (see Figure 1 in Paper I). 
\citet{Bertin77} had also reached a similar result for their 
three- and four-armed modes, and interpreted it based on the 
closeness of ILR and CR but with no mention of the increased 
likelihood of resonance overlapping. 

Having resonant zones of finite width for both the CR and ILR 
is an essential prerequisite for their subsequent competition. 
The $l$th Fourier component in the radial angle can trap stars 
into a growing resonant zone if $f_0(\textbf{\textit{J}})$ is
smooth and non-zero on a range of manifolds of constant 
$I_2=J_R-lJ_{\phi}/m$, and $\mu^{l}_m({\bf \Omega})$ can switch 
its sign in that neighborhood. Having some population of stars 
on circular orbits always guarantees the creation and growth 
of the CR zone because $I_2$ is identical to $J_R$ and 
$I_1=J_{\phi}/m$ can vary independently. However, if the ILR 
exists, its adiabatic invariant is defined by $I_2=J_R+J_{\phi}/m$ 
and the variation of $I_1$ in the vicinity of a resonant value 
implies the variation of $J_R$. Therefore, the ILR can develop 
a growing resonant zone if $f_0(\textbf{\textit{J}})$ has a 
functional dependence on both actions. Stellar disks with such 
a property will have non-zero radial velocity dispersions 
(warm disks). 

Let me now consider a group of stars moving on near-circular orbits 
that satisfy $\mu^{-1}_m({\bf \Omega})>0$ and have been suddenly 
trapped into resonance by the ILR. On the resonant tori, the orbits 
of those stars are deformed so that $I_2=J_R+J_{\phi}/m$ remains 
constant. This means that orbits are elongated as $J_R$ continuously 
increases from its tiny initial value and $J_{\phi}$ drops until 
$\mu^{-1}_m({\bf \Omega})$ becomes negative and phase space flows 
reach their turning points. At this moment, some stars are likely 
to find themselves on the integral manifolds of constant $J_R \gg 0$
and join the CR zone while they carry energy and angular momentum 
away from the ILR. A reverse migration is also possible. The 
switchings can only take place in the overlapping region of the ILR 
and CR where highly eccentric orbits live and 
$\mu^{-1}_m({\bf \Omega})<0<\mu^{0}_m({\bf \Omega})$. An effective 
early-stage overlap requires a non-zero $f_0(\textbf{\textit{J}})$ 
in the cross section ${\cal S}^{0}_{m} \cap {\cal S}^{-1}_{m}$ of 
the sets 
\begin{eqnarray}
{\cal S}^{0}_m &=& \left \{ {\bf\Omega} :  
\mu^{0}_m({\bf\Omega}) \approx{\cal O} 
         \left( \epsilon \right ) \right \}, \\
{\cal S}^{-1}_m &=& \left \{ {\bf\Omega} : 
\mu^{-1}_m({\bf\Omega})\approx{\cal O}
         \left( \epsilon \right ) 
\right \}.
\end{eqnarray}
As a result, infinitesimal wave amplitudes (in the linear regime)
can suppress their own growth due to the competition between
the CR and ILR. In the overlapping region 
${\cal S}^{0}_{m} \cap {\cal S}^{-1}_{m}$ neither $I_2=J_R$ 
(for $l=0$) nor $I_2=J_R+J_{\phi}/m$ (for $l=-1$) are conserved 
and that region is filled by chaotic orbits. The chaotic layer 
should be very narrow in the linear regime, but it will become 
thick in steady large-amplitude bars \citep{C83}.
 
The stability of warm stellar disks for $\Omega_p<\Omega_{\rm ILR}(m)$
can be deduced from an independent mathematical reasoning as follows. 
Equation (\ref{eq:growth-of-Delta-versus-s}) applies to any growing 
resonant gap in the phase space. Let me assume that the ILR and CR 
of a wave with the pattern speed $\Omega_p$ are widening their 
resonant zones with the rates 
\begin{equation}
\dot \Delta_{\rm CR}/\Delta_{\rm CR} \sim s_1/2,~~
\dot \Delta_{\rm ILR}/\Delta_{\rm ILR} \sim s_2/2.
\label{eq:Delta-for-CR-and-ILR}
\end{equation}
In an equilibrium disk with $d\Sigma_0/dR<0$, the probability 
of capture into the ILR is more than the CR, and according to 
(\ref{eq:Delta-for-CR-and-ILR}), we need $s_2>s_1$ to assure 
a faster expansion rate of the ILR zone. This inequality 
contradicts the basic assumption of linear analysis that 
separates the time-dependent part of perturbations as 
$\exp(-\imath \omega t)$ and assigns the same growing envelope 
and pattern speed to all parts of a normal mode. Therefore,
the only possibility for a disk with a monotonic density 
gradient will be $s_1=s_2=0$. In cold disks with 
$f_0(\textbf{\textit{J}})=\delta(J_R)g_0(J_{\phi})$,
where $\delta(J_R)$ is Dirac's delta function, the ILR
cannot open a resonant gap because it suffers from the
lack of fuel (stars) needed for a possible expansion 
along the curves of constant $I_2=J_R+J_{\phi}/m$. 
This automatically implies $s_2=0$ and a unique value 
is assigned to $s_1$. The CR zone can thus grow freely 
and destabilize the disk.
 
It is worth noting that with an inner cutout, the surface 
density rises in central regions, reaches to a maximum 
and then decays outwards. Denote $R_{\rm CR}$ and $R_{\rm ILR}$ 
as the corotation and inner Lindblad radii, respectively.
If the cutout is not sharp and the density peak lies fairly 
between $R_{\rm ILR}$ and $R_{\rm CR}$ 
(see the leftmost panel in Figure 8 of Evans \& Read 1998b), 
it will be possible to find two points of the same density 
on the curve of $\Sigma_0(R)$: one near the ILR on the 
rising part of $\Sigma_0(R)$ and the other near the CR 
on the falling side, i.e., 
$\Sigma_0(R_{\rm ILR})\approx \Sigma_0(R_{\rm CR})$.
Thus, the ILR and CR can manage to grow with the same 
rate $s_1=s_2$ and an unstable mode emerges. Such modes 
have been demonstarted in Figure 7 of \citet{ER98b} whose 
cutout functions immobilize the circular orbits of the 
galactic center and all near-radial orbits. Accordingly, 
the population of active stars in the overlapping region 
of the ILR and CR declines substantially and the simultaneous 
growth of both resonances becomes likely. By decreasing the 
pattern speed, the ILR moves outside the cutout radius and 
its resonant zone becomes visible to the CR. The stellar 
disk is therefore stabilized. The condition $s_1=s_2\not =0$ 
cannot be reached if $\Sigma_0(R_{\rm CR})$ differs 
remarkably from $\Sigma_0(R_{\rm ILR})$. This happens 
when both the ILR and CR lie outside the cutout radius or 
when the cutout is so sharp.

The above analysis shows how diverse the response of 
disk galaxies may be to non-axisymmetric excitations. 
\citet{T81} suggested that the two parameters $Q$ and $X$ 
decide the fate of non-axisymmetric density waves. 
Wave mechanics based on the behavior of resonances, 
shows more complex decisive factors that govern the 
evolution of density waves, specially near the critical 
values of $\Omega_p\approx \Omega_{\rm ILR}(m)$ and 
for non-monotonic gradients of equilibrium density.

\subsection{A Criterion for Global Stability}
\label{sec:criterion-for-stability}

I am now in a position to provide a simple criterion for the
stability of stellar disks in the absence of the ILR. It is the CR
that captures stars, contributes an angular momentum imbalance to
the stellar disk, and triggers unstable modes. Circumferential waves
will be {\it globally stable} if resonant gaps do not open near 
the CR for all possible pattern speeds. Consequently, $L_m(0)$ 
should remain zero, which according to 
equation (\ref{eq:torque-component-1}) implies
\begin{equation}
\Lambda^{m0}_{jk}=0,~~\forall j,k\ge 0.
\end{equation}
This condition is satisfied only if [see equations
(\ref{eq:define-Lambda}) and (\ref{eq:define-Phi-vs-Psi})]
\begin{equation}
\frac{\partial f_0}{\partial J_{\phi}}=0,~~ {\rm
for~all}~\textbf{\textit{J}} \in \Re^{+}\times \Re^{+}.
\label{eq:condition-stability-1}
\end{equation}
The stability criterion given in (\ref{eq:condition-stability-1})
can also be deduced from the earlier works of LBK and JH as I
explain below. An alternative form of (\ref{eq:total-torque}) is
given in equation (B6) of JH as
\begin{eqnarray}
{d{\cal L} \over dt} &=&
    -2ms\pi^2e^{2st} \sum_{l=-\infty}^{\infty}
    \int d\textbf{\textit{J}}
 \left(l{\partial f_0\over \partial J_R}
       +m{\partial f_0\over \partial J_{\phi}} \right)
       \nonumber \\
&{}& \times
{|\tilde V_{l}|^2 \over |l\Omega _R+m\Omega _{\phi} -\omega|^2}.
\label{eq:dL-dt-from-JH}
\end{eqnarray}
where
\begin{equation}
\tilde V_{l}=\sum_{j=0}^{\infty} \tilde a^m_j
\tilde \Psi^{ml}_j(\textbf{\textit{J}}).
\end{equation}
Dividing equation (\ref{eq:dL-dt-from-JH}) by $16\pi^4$ 
and changing the sign of $\omega$ lead to equation (28) in LBK.
The amount of angular momentum emitted/absrobed by the $l=0$ 
component is therefore determined by
\begin{equation}
\left [ {d{\cal L} \over dt} \right ]_{l=0}=-2(m\pi)^2
se^{2st} \int d\textbf{\textit{J}}
 {\partial f_0\over \partial J_{\phi}}
{|\tilde V_{0}|^2 \over |m\Omega _{\phi} -\omega|^2},
\label{eq:dL-dt-from-JH-CR}
\end{equation}
which reduces to equation (96) of \citet{GT79} as $s\rightarrow 0$.
It is seen that the sign of $\partial f_0/\partial J_{\phi}$
determines whether the angular momentum is added to, or drained from
the $l=0$ wave component. Furthermore, the CR will not imbalance 
${\cal L}$ if the condition (\ref{eq:condition-stability-1}) holds. 
This means that the condition (\ref{eq:condition-for-stationary-points}) 
cannot be violated according to my discussion of 
\S\ref{sec:origin-of-instabilities}, and $s$ should vanish for 
arbitrary perturbations. A stellar disk that satisfies
(\ref{eq:condition-stability-1}) will be hot because it is
impossible to reproduce the surface density gradient in the limit of
$J_R\rightarrow 0$ with a DF that does not depend on $J_{\phi}$. It
has now become more transparent why there is a correlation between
Toomre's (1964) $Q$ and the disk stability for non-axisymmetric
excitations. An illustrative example has been given in Figure 6 of
Paper I. Most of S-modes disappear when the parameter $N$ of the DF
is decreased and more stars are distributed on near-radial orbits.

\subsection{The Stability of Scale-Free Disks}
\label{sec:scale-free-disks}

Stability analysis of cuspy mass models has been a big puzzle in
stellar dynamics since \citet{Z76} computed the global modes of
Mestel's (1963) disk and showed the importance of $m=1$ excitations.
To tackle the singular center of Mestel's disk, he introduced
an inner cutout that freezes central stars with diverging orbital
frequencies. \citet{ER98a,ER98b} followed the same procedure but
for different cusp slopes. In spite of all these remarkable
efforts, what H02 illustrated for the frequency space of
scale-free potentials was indeed the key to resolve the
long-standing stability problem of scale-free disks without
any simplifying cutouts. The map of the angular sector that
H02 plotted in his Figure 1 for the scale-free potentials
\citep{TT97}
\begin{eqnarray}
V_0(R)= \biggl \{
\begin{array}{ll}
{\rm sign}(\beta)R^{\beta}, & ~ \beta \not =0,~
-1< \beta < 2, \\
\ln R, & ~ \beta=0,
\end{array}
\end{eqnarray}
will be an angular sector in the $(\Omega_R,\Omega_i)$-space
with the $\Omega_R$-axis and the straight line
\begin{equation}
\Omega_i=\left ( \frac{1}{\sqrt{2+\beta}}-\frac 12 \right )\Omega_R,
\end{equation}
being its lower and upper boundaries, respectively. The slope of the 
upper boundary is always less than or equal to $1/2$. It is therefore 
obvious that except for $m=1$, all $\mu^{-1}_m({\bf \Omega})=0$ and
$\mu^{0}_m({\bf \Omega})=0$ lines will intersect the frequency space 
(whatever $\Omega_p$ may be) and resonance overlapping between the CR 
and ILR can occur if there are enough number of non-circular orbits. 
In other words, the radial velocity dispersion $\tilde \sigma_u$
(defined in Evans \& Read 1998a) should exceed a critical value 
$\tilde \sigma_{u,\rm min}(m)$ to give rise to nonempty cross sections 
${\cal S}^{0}_{m}\cap {\cal S}^{-1}_{m}$. This is possible only 
for $m>1$ excitations and it suggests that scale-free disks are 
subject to $m=1$ instabilities. Finding $\tilde \sigma_{u,\rm min}(m)$
is an open, yet conceivable, problem that can generalize Toomre's 
(1964) criterion to non-axisymmetric perturbations of scale-free 
disks. I explained in \S\ref{sec:role-of-ILR} that the ILR cannot 
help to stabilize cold disks. That argument applies to self-consistent 
scale-free disks as well, and strongly justifies the existence
of $\tilde \sigma_{u,\rm min}(m)$ for $m>1$. Moreover, a stability 
condition like $\tilde \sigma_u> \tilde \sigma_{u,\rm min}(m)$ may 
correlate with Lynden-Bell's (1993, equation 6-23) criterion, which 
has been formulated based on the angular dispersion of lobe tumble 
rates.

\section{DISCUSSION AND CONCLUSIONS}
\label{sec:discussion-and-conclusions}

I had already given an evidence in \S\ref{sec:modes-in-frequency-space} 
that LBK's theory cannot be extended to large growth rates. Here I provide
a more detailed analysis of LBK's mode mechanism. Taking the $s\rightarrow 0$ 
limit of (\ref{eq:dL-dt-from-JH}) leads to equation (30) of LBK. 
The resulting integrand of the $l$th component involves the term 
$\delta(l\Omega_R+m\Omega_{\phi}-\omega)$ but its 
argument never becomes zero for $l\le -1$ as long as $\Omega_p$ 
exceeds $\Omega_{\rm ILR}(m)$. Once the delta function vanishes for 
$\Omega_p>\Omega_{\rm ILR}(m)$, the angular momentum content of $l\le -1$ 
components remains zero. Therefore, angular momentum transfer between 
inner and outer resonances as suggested by LBK, is feasible only for 
$\Omega_p < \Omega_{\rm ILR}(m)$. This is a constraint on the 
applicability of LBK's theory.

According to the arguments of \S\ref{sec:role-of-ILR} and the WKB 
theory, the ILR is not transparent to short wavelength disturbances 
with $X\gg 1$ \citep{BT08,Mark74} and any developing spiral structure 
must be damped unless the ILR remains invisible to the CR. Therefore, 
an inconsistent point in LBK's paper is the imagination of an 
angular-momentum-transferring spiral structure while the condition 
$\Omega_p <\Omega_{\rm ILR}(m)$ has already abandoned the existence 
of such structures. An issue yet needs to be explained: If there is 
no spiral structure in stable disks, which mechanism does transfer 
the angular momentum between inner and outer resonances? 
In fact, in the limit of $s\rightarrow 0$ 
the density components $\Sigma^{l-}_1$ and $\Sigma^{l+}_1$ become 
stationary waves that are azimuthally separated by a phase shift of 
$90^{\circ}$ (like the components of mode B1 in 
Figure \ref{pic:components-modes-B1-S2}). Thus, they exert 
opposite gravitational torques on each other without any need 
for a communicating spiral structure.

In this paper I decomposed unstable modes of a model disk galaxy to
its constituent Fourier components and showed how different
components experience a gravitational torque. My results clearly
showed that only the Fourier component associated with the CR
generates a resonant zone in the phase space, and other components
exchange angular momentum far from resonances. This result led me 
to introduce a new dynamical mechanism that triggers unstable 
modes. According to my calculations, an irreversible resonant 
capture of stars into the CR causes a synchronous precession of 
their orbital axes, which in turn, support a rotating density 
pattern. The emerged pattern grows because the resonant zone 
expands in the frequency space.

The irreversibility of resonant trapping is the most destructive
event that happens in a stellar disk when a group of resonant stars with
$d_i<I_1<d_o$ gain angular momentum forever. An immediate consequence 
of this phenomenon is the angular momentum imbalance in the disk, 
which requires proper reaction of other stars in order to conserve 
${\cal L}$. This is how the Fourier components of $l\not =0$ are 
generated and the angular momentum transfer between them is initiated. 
Reacting stars do not have a librating $\theta$ and they cooperate 
in a way that the quantity $\vert \mu^0_m({\bf \Omega})\vert$ remains 
small for all components.

Using the maps of resonance zones in the phase and frequency spaces, 
I argued that resonance overlapping should stop the growth of unstable 
modes and stabilize stellar disks in the presence of an ILR. I showed 
that for $m>1$ the emergence of an ILR is inevitable in scale-free 
models. The competition between the CR and ILR can therefore stabilize
sufficiently warm scale-free disks against $m>1$ excitations. 
\citet{ER98b} suggested in \S6.1 of their paper that self-consistent 
scale-free disks (without inner cutouts) do not admit growing 
non-axisymmetric modes at all, and they ruled out the possibility of 
a critical temperature. Their prediction is in agreement with my results 
except in two aspects: (i) Since the line $\mu^{-1}_{1}({\bf \Omega})=0$ 
does not intersect the frequency space of scale-free models, there is 
no ILR to compete with a growing CR and density waves will be amplified
for $m=1$. (ii) The ILR in cold disks is very special and it cannot 
develop a resonant zone of finite size. It is therefore hard to believe 
a serious influence by the ILR on the phase space flows up to a critical
temperature.

Note that each unstable mode dominates an isolated region of the 
frequency (action) space until a chaotic layer emerges due to resonance 
overlapping. The rate by which the chaotic layer diffuses itself 
in the phase space, is determined by the resistance of KAM tori 
against competing resonant zones. The limited space that the fastest 
growing bar mode occupies after its saturation (Khoperskov et al. 2007), 
is an $N$-body evidence for such a resistance of spiral modes that 
fill outer regions of the cored exponential disk.

The other implication of resonance overlapping, which has
observational support too, is that the central bar in grand-design
barred spiral galaxies must join the spiral arms at the tips of the
bar, which are the overlapping regions of two neighboring B and S
modes in the frequency space (see Figure \ref{pic:zones-overlap}). I
note that the pattern speeds of the spiral and bar components are
not the same because these structures are associated with different
modes. The overlapping region is filled by chaotic orbits that yield
the turbulent mixing of density waves. This process can enhance star
formation near the tips of the bar. If we accept this scenario,
barred spiral galaxies should have been formed from hot stellar
disks whose eigenfrequency spectra include few spiral modes. On the
other hand, flocculent spirals with many wave packets will be the
natural destiny of initially cold disks that give birth to a rich
family of spiral modes.

The mathematical background for the nonlinear evolution of modes was
developed in Paper I but that formulation is valid as long as
averaging is allowed over angle variables. In the presence of
chaotic orbits it is impossible to average out resonant angles and
some modifications are required to deal with the CBE in its full
nonlinear form. I will present such modifications in Paper III with
the aim of discovering the dynamical processes that generate
different classes of barred and spiral structures.

\acknowledgments

I thank Scott Tremaine and Chris Hunter for their encouragement
and stimulating discussions in the course of my calculations and
writing the manuscript. I also express my sincere thanks to the 
referee whose illuminating comments helped me to further my 
calculations and thoroughly revise the presentation and 
arguments of sections 5 through 8. The major part of this 
research was supported by the Institute for Advanced Study 
at Princeton, where I was an astrophysics visitor.

\end{document}